\documentclass[12pt]{article}
\usepackage{epsfig,picinpar,floatflt,amssymb}
\newcommand{\resection}[1]{\setcounter{equation}{0}\section{#1}}

\def\goto{\longrightarrow}

\def\bd{\begin{displaystyle}}
\def\ed{\end{displaystyle}}
\def\ba{\begin{array}}

\def\ea{\end{array}}
\def\EQ{\begin{equation}}
\def\EN{\end{equation}}
\def\bea{\begin{eqnarray}}
\def\eea{\end{eqnarray}}
\def\beano{\begin{eqnarray*}}
\def\eeano{\end{eqnarray*}}

\def\lab{\label}

\begin{document}
\oddsidemargin 5mm
\setcounter{page}{0}
\newpage     
\setcounter{page}{0}
\begin{titlepage}
\begin{flushright}
SISSA/EP/41/2007
\end{flushright}
\vspace{0.5cm}
\begin{center}
{\large {\bf Kink Confinement and Supersymmetry}} \\
\vspace{1.5cm}
{\bf Giuseppe Mussardo} \\
\vspace{0.8cm}
{\em International School for Advanced Studies, Via Beirut 2-4, 
34013 Trieste, Italy} \\ 
{\em Istituto Nazionale di Fisica Nucleare, Sezione di Trieste}\\
\end{center}
\vspace{6mm}
\begin{abstract}
\noindent
We analyze non-integrable deformations of two-dimensional $N=1$ supersymmetric quantum field theories with kink excitations. As example, we consider the multi-frequency Super Sine Gordon model. At weak coupling,  this model is robust with respect to kink confinement phenomena, in contrast to the purely bosonic case. If we vary the coupling, the model presents a sequence of phase transitions, where pairs of kinks disappear from the spectrum. The phase transitions fall into two classes: the first presents the critical behaviors of the Tricritical Ising model, the second instead those of the gaussian model. In the first case, close to the critical point, the model has metastable vacua, with a spontaneously supersymmetry breaking. When the life-time of the metastable vacua is sufficiently long, the role of goldstino is given by the massless Majorana fermion of the Ising model. On the contrary, supersymmetry remains exact in the phase transition of the second type.  
\end{abstract}
\vspace{5mm}
\end{titlepage}
\newpage

\setcounter{footnote}{0}
\renewcommand{\thefootnote}{\arabic{footnote}}

\resection{Introduction}
In two dimensions there are several bosonic field theories that have kinks as the basic excitations of their spectrum, the simplest example being  the Ising model in its low temperature phase (see, for instance, \cite{Fradkin}). Other examples are provided, for instance, by the minimal models 
${\cal M}_m$ ($m=3,4,\ldots)$ of Conformal Field Theory perturbed by the operator $\Phi_{1,3}$ \cite{phi13}: these are off-critical theories with $(m-1)$ degenerate vacua, connected by massive kinks.

\vspace{5mm}

\begin{figure}[ht]
\hspace{45mm}
\vspace{10mm}
\psfig{figure=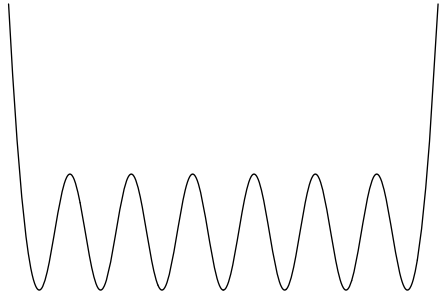,height=4cm,width=6cm}
\vspace{1mm}
\caption{{\em Effective potential $U(\varphi)$ of a quantum field theory 
with kink excitations.}}
\label{potential}
\end{figure}

\noindent
At the lagrangian level, theories with kink excitations are described by a scalar real field $\varphi(x)$ with an action ${\cal A} = \int d^2x\, {\cal L}$ and a Lagrangian density 
\EQ
{\cal L}\,=\,\frac{1}{2} \left(\partial_{\mu} \,\varphi\right)^2 - 
U(\varphi) \,\,\,,
\EN 
where the potential $U(\varphi)$ possesses a set of degenerate minima at $\varphi_k^{(0)}$ ($k=1,2,\ldots,n$), with $\varphi_{k-1}^{(0)} < \varphi_k^{(0)} < \varphi_{k+1}^{(0)}$. While these minima correspond to the different vacuum states $\mid 0 \,\rangle_k$ of the associate quantum field theory, the kink states\footnote{The quantity $\theta$ in the symbol $\mid K_{kl}(\theta)\,\rangle$ is the rapidity 
variable of the kinks, parameterising their relativistic dispersion relation $E = M_{kl}\,\cosh\theta$, $P = M_{kl} \,\sinh\theta$, where $M_{kl}$ is their mass.} $\mid K_{kl}\,(\theta)\,\rangle$ can be put in correspondence with the static solutions of the classical equation of motion 
\EQ
\partial^2_{x}\,\varphi(x) \,=\,U'\left[\varphi(x)\right]\,\,\,,
\EN 
with boundary conditions $\varphi(-\infty) \,=\,\varphi_k^{(0)}$ and $\varphi(+\infty) = \varphi_l^{(0)}$, where $l = k \pm 1$. The semi-classical quantization of these solutions has been discussed in \cite{DHN,GJ,Mus}.  Conventionally $\mid K_{k,k+1}(\theta) \,\rangle$ denotes the {\it kink} state between the pair of vacua $\mid 0 \,\rangle_k$, $\mid 0 \,\rangle_{k+1}$ while the corresponding {\it anti-kink} is 
associated to $\mid K_{k+1,k}(\theta)\,\rangle$.

A popular example of such a lagrangian theory is the Sine-Gordon model, described by the potential 
\EQ
U(\varphi) \,=\,-\mu \,\cos\alpha\varphi\,\,\,, 
\EN   
that has an infinite number of degenerate vacua at $\varphi_n^{(0)} =  2\pi n/\alpha$. The classical kink configurations that interpolate between two nearest ones are given by 
\EQ
\varphi_{\pm} (x) \,=\,\frac{4}{\alpha} \arctan 
[\exp(\pm \,\alpha\,\sqrt{\mu}\,x)] \,\, , 
\label{soliton}
\EN 
($+$ refers to the kink whereas $-$ to the antikink), so that, going from $x =-\infty$ to $x=+\infty$, the field has a jump $\pm 2\pi/\alpha$.  
 
The kinks obviously owe their existencies to the $n$-fold degenerate vacuum structure of the theory. However, such a degeneracy is often a fragile condition, which can be easily broken by inserting additional operators into  the action. For instance, by adding an extra trigonometric interaction 
$\lambda\,\cos\omega\varphi$ (with $\omega \neq \alpha$) to the Sine-Gordon model, the landscape of its potential changes from the situation (a) to the situation (b) of Figure 2, even for small value of the coupling constant $\lambda$. The unbalance of the vacua has a drastic consequence on the particle 
content of the theory: once the degeneracy of the original minima is lifted, some kink excitations (if not all) disappear from the spectrum of the perturbed theory, i.e. they get confined. As a consequence, the linear confinement potential between the kink and antikink causes the collapse of this pair into a string of bound states \cite{Mccoy-Wu,DMS,DM,FonsecaZam,Rutskevic} (see Figure 3).

\vspace{3mm}
\begin{figure}[h]\label{SGfigure}
\hskip 5pt
\begin{minipage}[b]{.45\linewidth}
%\centering\psfig{figure=m1.eps,width=\linewidth}
\centering\psfig{figure=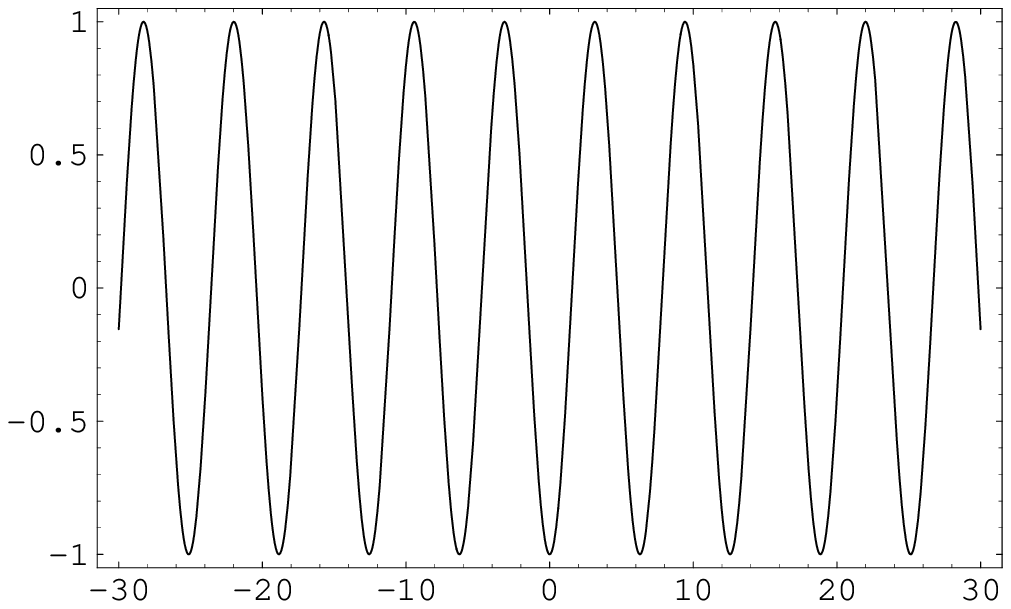,width=\linewidth}
%\caption{(a)}
%\vspace{1mm}
\begin{center}
{\bf (a)}
\end{center}
\end{minipage} \hskip 30pt
\begin{minipage}[b]{.45\linewidth}
\centering\psfig{figure=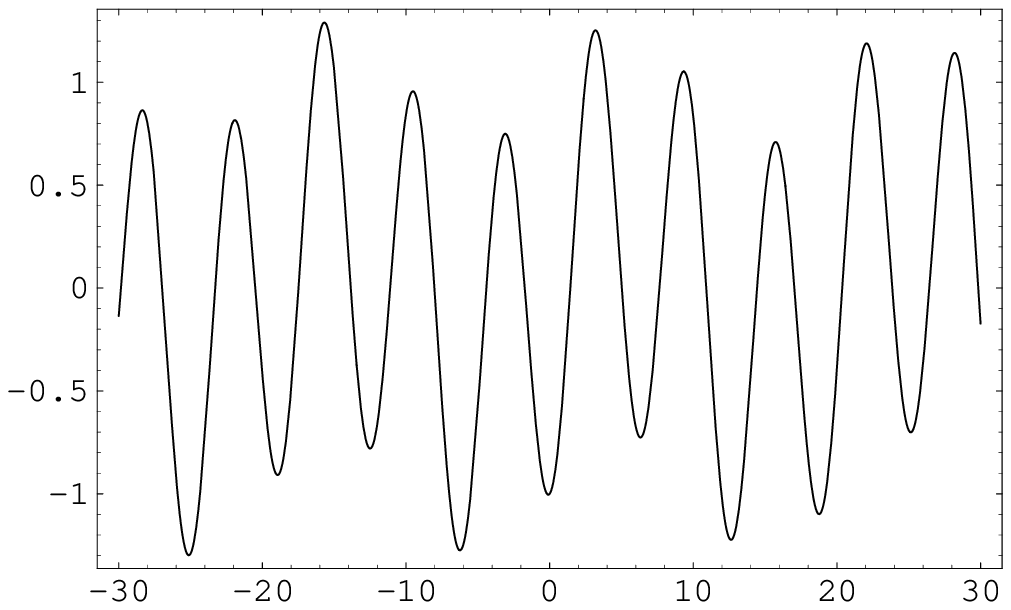,width=\linewidth}
%\centering\psfig{figure=m2.eps,width=\linewidth}
%\caption{(b)}
%\vspace{1mm}
\begin{center}
{\bf (b)}
\end{center}
\end{minipage}
\caption{{\em Potential of (a) Sine-Gordon model; (b) Perturbed 
Sine-Gordon model.}}
%\vspace{5mm}
%\begin{center}
%{\bf Figure 2} 
%\end{center}
\end{figure}

Apart some special cases (some of them discussed in Section \ref{multiSG}), the above situation seems to be the general scenario for  the bosonic theories with kink excitations. The situation changes, however, if we add fermions. In fact, as shown below, adding fermions to a bosonic theory,  while requiring a supersymmetry of the corresponding action, has the effect to stabilize the vacuum states. In other words, supersymmetric theories are less sensitive to the confinement phenomena of the kinks, if 
put under a weak coupling deformation of the original action. At strong coupling, instead, they may present interesting phenomena of metastable vacuum states, which signal certain phase transitions. The 
two situations can be related to the spontaneous breaking of the supersymmetry that occurs at certain vacua.

The paper is organised as follows. In the next section we briefly summarise the confinement phenomena of kinks in pure bosonic theories. In section 3 we introduce the supersymmetric theories and discuss 
their conformal limit. Section 4 deals with the Super Sine Gordon model, i.e. an integrable quantum field with kink excitations. In Section 5 we address the stability of the kinks under the supersymmetric deformation of the action that leads to the multi-frequency Sine Gordon model. In Section 6 we discuss the phenomena of meta-stable states that occur at finite value of the coupling constant of the perturbing operator. Our conclusions are presented in Section 7.

\vspace{3mm}
\begin{figure}[h]\label{linearfigure}
\hskip 30pt
\begin{minipage}[b]{.45\linewidth}
%\centering\psfig{figure=m1.eps,width=\linewidth}
\vspace{5mm}
\centering\psfig{figure=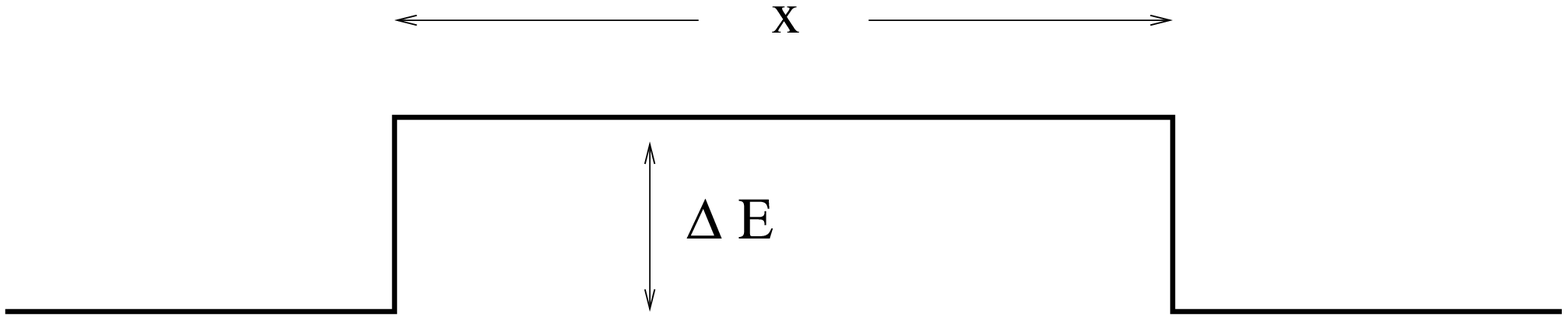,width=\linewidth}
%\caption{(a)}
%\vspace{1mm}
\begin{center}
{\bf (a)}
\end{center}
\end{minipage} \hskip 55pt
\begin{minipage}[b]{.25\linewidth}
\centering\psfig{figure=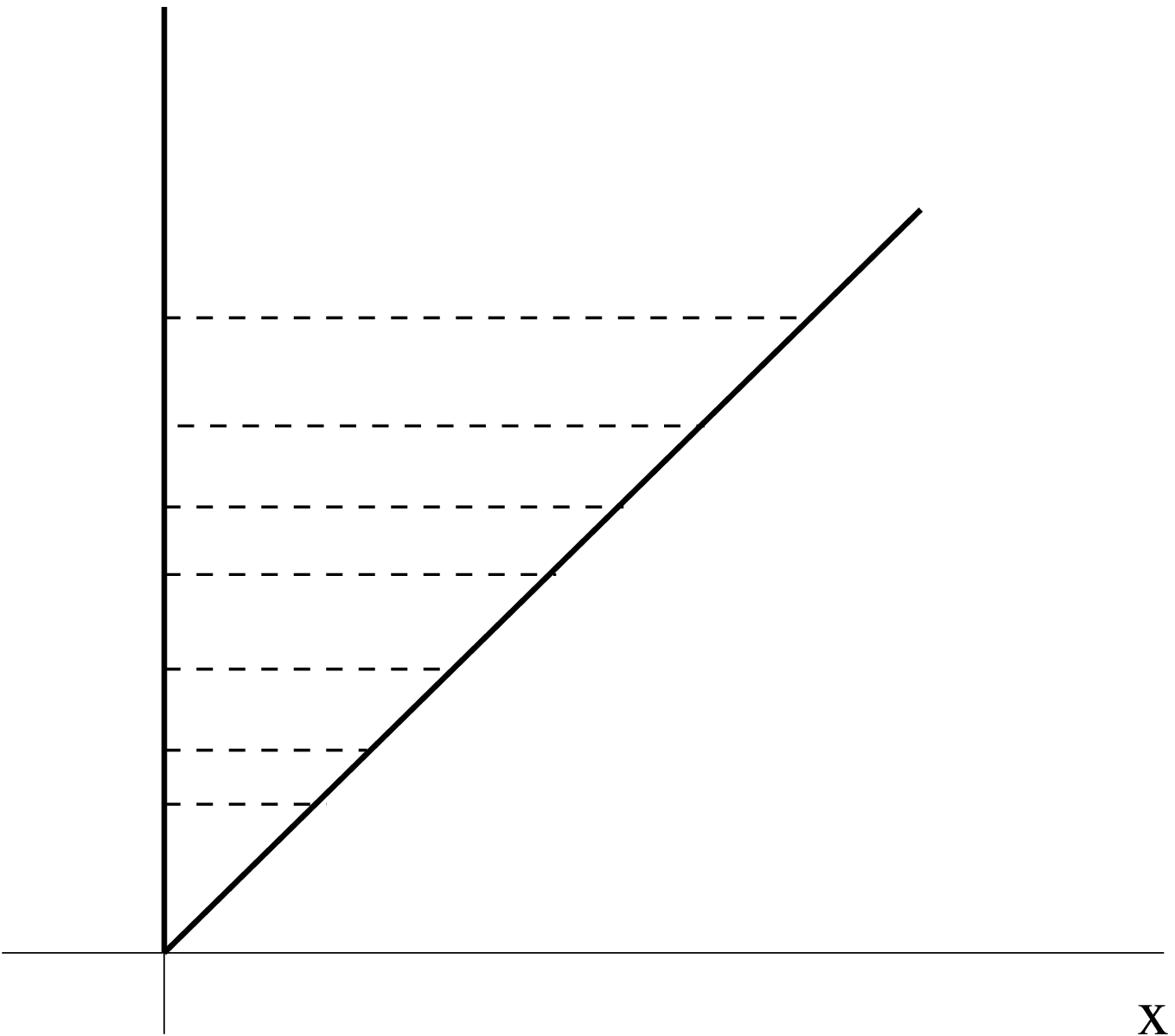,width=\linewidth}
%\centering\psfig{figure=m2.eps,width=\linewidth}
%\caption{(b)}
%\vspace{1mm}
\begin{center}
{\bf (b)}
\end{center}
\end{minipage}
\caption{{\em (a) Kink-antikink state at distance $x$, where $\Delta E$ is the gap  of the unbalanced vacua.  The tendency of the system to shrink, in order to minimise the energy of this configuration, 
gives rise to the linear potential (b), with slope given by $\Delta E$. The dashed lines are the bound states of the kink-antikink pair.}}
%\vspace{5mm}
%\begin{center}
%{\bf Figure 2} 
%\end{center}
\end{figure}

\resection{Kink Confinement in Bosonic Theories}\label{residueconf}

Quantum field theories with kink excitations are not necessarily integrable: $\varphi^4$ theory in its $Z_2$ broken phase, for instance, is not an integrable model although it has kink excitations. At the same time, the confinement of the kinks is also a general phenomenon,  i.e. it occurs in deformation of both integrable and non-integrable theories.  In order to have a certain analytic control of this phenomenon, it is  however more convenient to discuss the confinement of the kink excitations  that takes place in perturbed integrable theories. The reason is that, for these theories, we can rely on a well-defined perturbation approach, the so called Form Factor Perturbation Theory  (FFPT) \cite{DMS,DM}, that allows us to analitically follow the fate of the kink states: properly interpreted, the results obtained in the context of perturbed integrable theories turn out to be useful also to understand the confinement phenomena of a generic model. 

Based on what stated above, let us consider an integrable theory with purely bosonic degrees of freedom described by the action ${\cal A}_0$. We assume that such a theory has $n$ degenerate vacua $\mid 0 \rangle_k$, ($k = 1,2,\ldots, n$) and therefore its spectrum contains the topological excitations $\mid K_{k,k\pm 1}(\theta) \rangle$ (the theory may also have bound states thereof). All kinks have the same mass $M$. 

Suppose now that we perturb the system using a field $\Upsilon(x)$: does the perturbed action 
\EQ
{\cal A}\,=\,{\cal A}_0 + \lambda\,\int \,d^2 x\,\,\Upsilon(x) \,\,\,,
\label{perturbedaction}
\EN
still support the topological kinks $\mid K_{k,k\pm 1}(\theta)\,\rangle$ as asymptotic states? As shown in \cite{DMS,DM}, to answer this question is sufficient to compute the correction $\delta M_{kl}$ to the mass of the kinks due to the field $\Phi(x)$. At the first order in $\lambda$, this correction is given by 
\EQ
\delta M_{kl}^2 \,\simeq \,2\,\lambda \,F_{kl}^{\Upsilon}(i\pi)\,\,,
\label{deltam}
\EN
where 
\EQ
F_{kl}^{\Upsilon}(\theta) \,\equiv\,_k\langle 0|\Upsilon(0)|K_{kl}(\theta_1)
\,K_{lk}(\theta_2)\rangle\,\, 
\label{aabar}
\EN
is the kink-antikink Form Factor of the operator $\Upsilon(x)$, with $\theta = \theta_1 - \theta_2$. In integrable field theories (and this is the important technical point for which it is useful to consider these theories), Form Factors of a generic scalar operator ${\cal O}(x)$ can be computed exactly. In fact, they satisfy manageable functional equations in virtue of the simple form assumed by the unitarity and crossing symmetry equations 
\cite{KW,Smirnov}. 

Consider the two--particle case: shortening the notation and denoting the kink state $\mid K_{k,k+1}\rangle$ by $a$ and the anti-kink state $\mid K_{k+1,k}\rangle$ by $\bar a$, for the Form Factor of the scalar operator ${\cal O}$ we have the following equations 
\EQ 
F_{a\bar{a}}^{\cal O}(\theta)\,=\,S_{a\bar{a}}^{b\bar{b}}(\theta)\,
F_{\bar{b}b}^{\cal O}(-\theta)\,\,,
\label{uni}
\EN
\EQ
F_{a\bar{a}}^{\cal O}(\theta+2i\pi)\,=\, e^{-2i\pi\gamma_{{\cal O},a}}\,
F_{\bar{a}a}^{\cal O}(-\theta)\,\,.
\lab{cross}
\EN
In the first of these equations the sum on $b$ is on all kink states which start from the vacuum $\mid 0\,\rangle_k$, i.e. it includes the two states $\mid K_{k,k\pm 1} \rangle$. It expresses the fact that in an 
integrable theory the two-particle threshold is the only unitarity branch point in the plane of the Mandelstam variable $s=(p_a+p_{\bar{a}})^2 = 4 M^2 \cosh^2(\theta/2)$, the discontinuity across the cut being determined by the two-body scattering amplitude $S_{a\bar{a}}^{b\bar{b}}$. 

In the second equation the explicit phase factor $e^{-2i\pi \gamma_{{\cal O},a}}$ is inserted to take into account a possible semi-locality of the kink with respect to the operator ${\cal O}(x)$ \footnote{Consistency of eq.\,(\ref{cross}) requires $\gamma_{{\cal O}, \bar{a}} = - \gamma_{{\cal O},a}$.}. When $\gamma_{{\cal O},a}  =0$, there is no crossing symmetric counterpart to the unitarity cut but when $\gamma_{{\cal O},a}\neq 0$, there is a non-locality discontinuity in the $s$-plane with $s=0$ as branch point. However, in the rapidity parameterisation, there is no cut because the different Riemann sheets of 
the $s$-plane are mapped onto different sections of the $\theta$-plane; the branch point $s=0$ is mapped onto the points $\theta=\pm i\pi$ which become,  in this way, the locations of simple {\em annihilation} poles, whose residues are given by \cite{Smirnov} (see also \cite{Yurov})  
\EQ
-i\,{\mbox Res}_{\,\theta=\pm i\pi}\,F_{a\bar{a}}^{\cal O}(\theta)\,=\,
(1-e^{\mp 2i\pi\gamma_{{\cal O},a}})\,\,_k\langle 0|{\cal O}|0\rangle_k
\,\,.
\label{pole}
\EN
Notice that these residues are expressed by the product of two terms: the first, determined by the semi-local index, while the second given by the vacuum expectation value of the field ${\cal O}(x)$.  

With the above information, let's come back to the perturbed action (\ref{perturbedaction}). It is easy to see that there is a drastic effect on the kink spectrum if (a)  the perturbing operator $\Upsilon(x)$ is 
semi-local with respect to them and (b) if its expectation value of the vacuum $\mid 0 \,\rangle_k$ is different from zero. 

If both conditions are fulfilled, the application of eq.\,(\ref{deltam}) leads to an infinite correction to the 
kink masses. This divergence has to be interpreted, in turn, as the technical manifestation of the fact that the kink starting from  the vacuum $\mid 0\,\rangle_k$ no longer survive as asymptotic particles of 
the perturbed theory. In order to clarify better this point, it may be useful to discuss a couple of 
examples\footnote{It may be also useful to see \cite{DG}.} . 

%\vspace{3mm}

%\noindent
%{\em {\bf Low-temperature Ising model in a magnetic field}}
\subsection{Low-temperature Ising model in a magnetic field}

The simplest example of the above scenario is given by the Ising model. In its low-temperature phase it has two degenerate vacua, connected by a kink and an anti-kink excitation. Suppose now this system is coupled to an external magnetic field $h$, so that the perturbed action is given by (${\cal A}_0$ is the integrable action of the low-temperature Ising) 
\EQ
{\cal A}\,=\,{\cal A}_0 + h\,\int\,d^2x\,\sigma(x) \,\,\,.
\label{isingmagneticfield}
\EN 
The magnetization operator $\sigma(x)$ is non-local with respect to the kinks, with non-local index given by $\gamma_{\sigma,a} = \frac{1}{2}$ \cite{KadanoffCeva,Yurov}.  Hence, the two-particle Form Factor of the operator $\sigma(x)$ should have a pole at $\theta = \pm i \pi$. This is confirmed by the exact expression of this matrix element, given (up to normalization) by \cite{Yurov}  
\EQ
\langle 0 \mid \sigma(0)\mid \,a(\theta_1)\,\bar{a}(\theta_2) \,\rangle \,=\,
\tanh\frac{\theta_1-\theta_2}{2} \,\,\,.
\EN 
Consequently, the kinks disappear from the spectrum of the perturbed theory (\ref{isingmagneticfield}), producing though an infinite  tower of bound states\footnote{ Not all of them are stable: the number of the stable ones changes by moving in the plane with axes ($h,T$) \cite{FonsecaZam,DGM}: near the magnetic axis, the decay processes of the unstable particles can be computed by using the Form Factor 
Perturbation Theory \cite{DGM}.} \cite{Mccoy-Wu,DMS,DM,FonsecaZam,Rutskevic}. The confinement of the kink is evident from Figure 4: no matter how small the magnetic field $h$ may be, the two original vacua are no longer degenerate in the perturbed theory. Hence, the perturbed theory 
(\ref{isingmagneticfield}) does not have any longer kink excitations. 

\vspace{10mm}
\begin{figure}[h]
\hskip 40pt
\begin{minipage}[b]{.35\linewidth}
\centering\psfig{figure=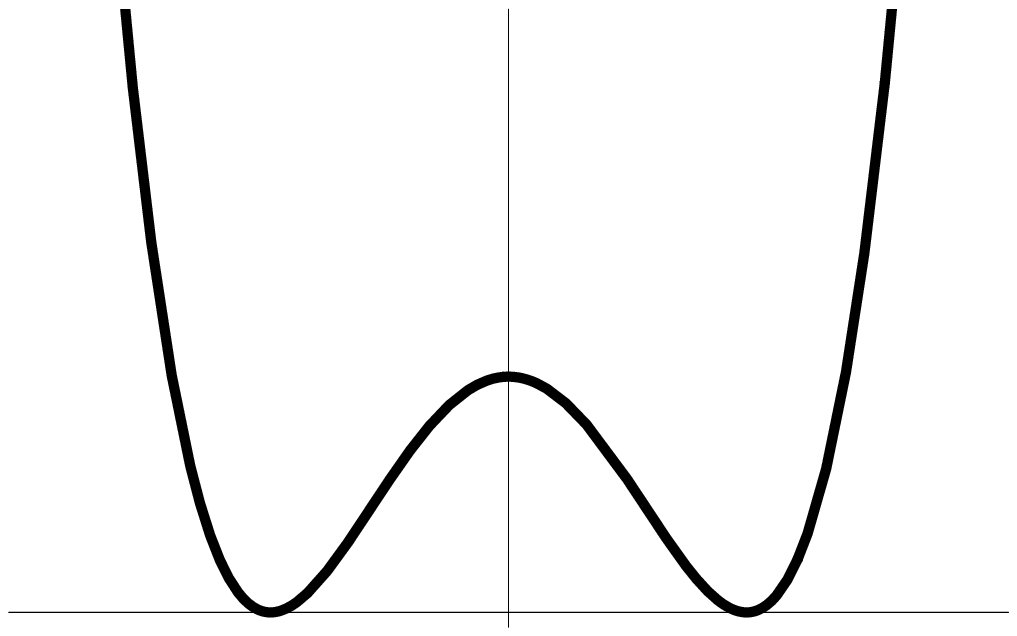,width=\linewidth}
%\caption{(a)}
%\vspace{1mm}
\begin{center}
{\bf (a)}
\end{center}
\end{minipage} \hskip 30pt
\begin{minipage}[b]{.35\linewidth}
\centering\psfig{figure=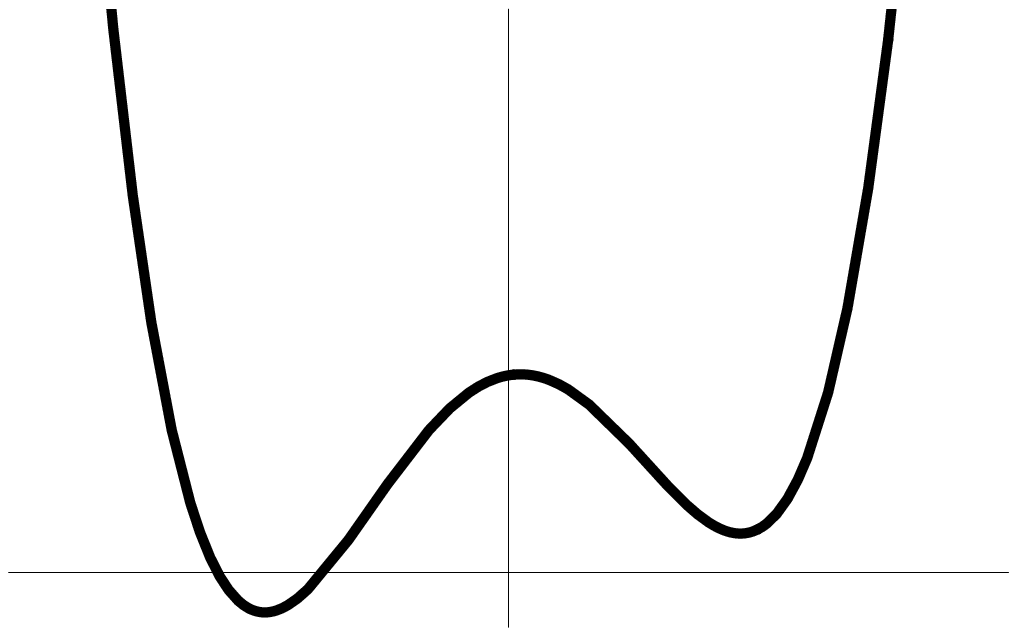,width=\linewidth}
%\caption{(b)}
%\vspace{1mm}
\begin{center}
{\bf (b)}
\end{center}
\end{minipage}
\caption{{\em Effective potential of the Ising model in (a) low-temperature 
phase and (b) in the presence of a magnetic field.}}
%\vspace{5mm}
%\begin{center}
%{\bf Figure 2} 
%\end{center}
\end{figure}

\vspace{3mm}

%\noindent
%{\em {\bf Multi-frequency Sine-Gordon model}}
\subsection{Multi-frequency Sine-Gordon model}\label{multiSG}
%\noindent
As a second example, let's consider the perturbed Sine-Gordon 
model
\EQ
{\cal L} \,= \,
\frac{1}{2}(\partial_\mu\varphi)^2 + \mu\,\cos\alpha\varphi +
\lambda\cos(\beta\varphi+\delta) \,\,\,.
\label{lagrangian}
\EN
This system has been analysed in detail in \cite{DM}, so here we briefly remind some of the main results. First of all, the quantum model does not depend separately by $\mu$ and $\lambda$, it rather depends on the dimensionless variable 
\[
\eta \equiv \lambda\mu^{-(1-\Delta_\beta)/(1-\Delta_\alpha)} =
\lambda\mu^{-(8\pi-\beta^2)/(8\pi-\alpha^2)}\,\, .
\label{rg}
\]
In terms of $\eta$, we can identify two perturbative regimes of the model (\ref{lagrangian}) where it is possible to apply the FFPT: the first regime is obtained in the limit $\eta \rightarrow 0$, while the other is reached for $\eta \rightarrow \infty$. In the last case, one has simply to swap the role played 
by the two operators.  

Let us consider, for instance,  the system in the vicinity of $\eta = 0$. In the unperturbed situation ($\eta = 0$), the field undergoes to the following discontinuity $\varphi \goto \varphi + 2\pi/\alpha$ across a soliton configuration. As a consequence, the exponential operator $e^{i\beta\varphi(x)}$ is semi-local with respect to the kink with semi-locality index $\gamma_{\alpha,s} = \beta/\alpha$. According to 
eq.(\ref{pole}), the kink-antikink Form Factor of the perturbing operator $\Upsilon = \cos(\beta\varphi + \delta)$ contains annihilation poles at $\theta=\pm i\pi$, with residues given by \footnote{We choose as a 
vacuum state of the unperturbed theory the one at the origin, characterised by $\langle 0|\varphi|0\rangle = 0$. With this choice the unperturbed theory is invariant under the reflection $\varphi\goto -\varphi$ and therefore it holds the equality $\langle 0| e^{i\epsilon\varphi}|0\rangle =
\langle 0|e^{-i\epsilon\varphi}|0\rangle$.} 
\EQ
-i\,{\mbox Res}_{\,\theta=\pm i\pi}F_{a\bar{a}}^\Phi(\theta) \,=\, 
\left[\cos\delta-\cos(\delta\mp 2\pi\beta/\alpha)\right]\,
\langle 0|e^{i\beta\varphi}|0\rangle\,\,\,.
\label{respsi}
\EN
Hence, for generic values of the frequency $\beta$ (and the phase-shift $\delta$) of the perturbing operator $\Upsilon(x)$, the above residues are different from zero. Consequently, the kink and the anti-kink of the original Sine-Gordon model become unstable excitations of the perturbed theory. This result 
that can be easily understood by noting that the perturbed lagrangian (\ref{lagrangian}) loses its 
original $2\pi/\alpha$-periodicity (see Figure 2).

The possibility to control the changes of the spectrum in both perturbative limits ($\eta \rightarrow 0$ and $\eta \rightarrow \infty$) allows one to deduce interesting information about its evolution in the intermediate, non-perturbative region. In certain cases, for instance, there may be a topological 
excitation in one limit which is no longer present in the other. When this happens, the very nature of topological excitations requires that a change in the vacuum structure of the theory takes place 
somewhere in the non-perturbative region, namely that a quantum phase transition occurs.  Lines of phase transition are then expected to appear in the multi--frequency SG model for particular values of the parameters: as discussed in \cite{DM}, this circumstance occurs when $|\delta| = \pi/n$ and  $\beta/\alpha = 1/n$. Once these values of the parameters are fixed, by varying $\eta$ a change of the vacuum state takes place at $\eta_c$: at this critical value, the quantum field theory may be regarded as associated to a massless Renormalization Group flow that interpolates between the conformal field theory with central charge  $c=1$ (in the ultravioled regime) and the one associated to the Ising model, with $c = 1/2$ (in the infrared regime). The simplest way to see this, is to note  that, at $\eta_c$, the effective theory at the vacuum nearby the origin is given by the massless $\phi^4$ associated to the class of universality of the Ising model: for $\eta $ slightly smaller than $\eta_c$, there are two degenerate ground states separated by a potential barrier, for $\eta$ slightly greater than $\eta_c$ there is only one vacuum with a parabolic shape, while at $\eta=\eta_c$  the quadratic term vanishes, giving rise to an effective  $\phi^4$ behaviour with massless excitations (see Figure 5) 

\vspace{10mm}
\begin{figure}[h]
\hskip 35pt
\begin{minipage}[b]{.25\linewidth}
\centering\psfig{figure=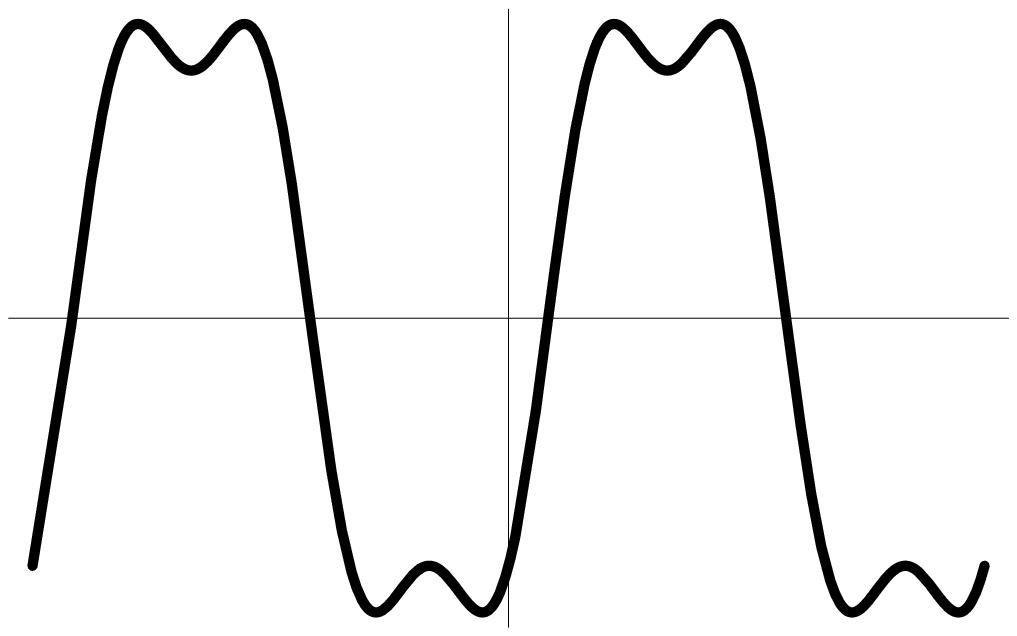,width=\linewidth}
%\caption{(a)}
%\vspace{1mm}
\begin{center}
{\bf (a)}
\end{center}
\end{minipage} \hskip 30pt
\begin{minipage}[b]{.25\linewidth}
\centering\psfig{figure=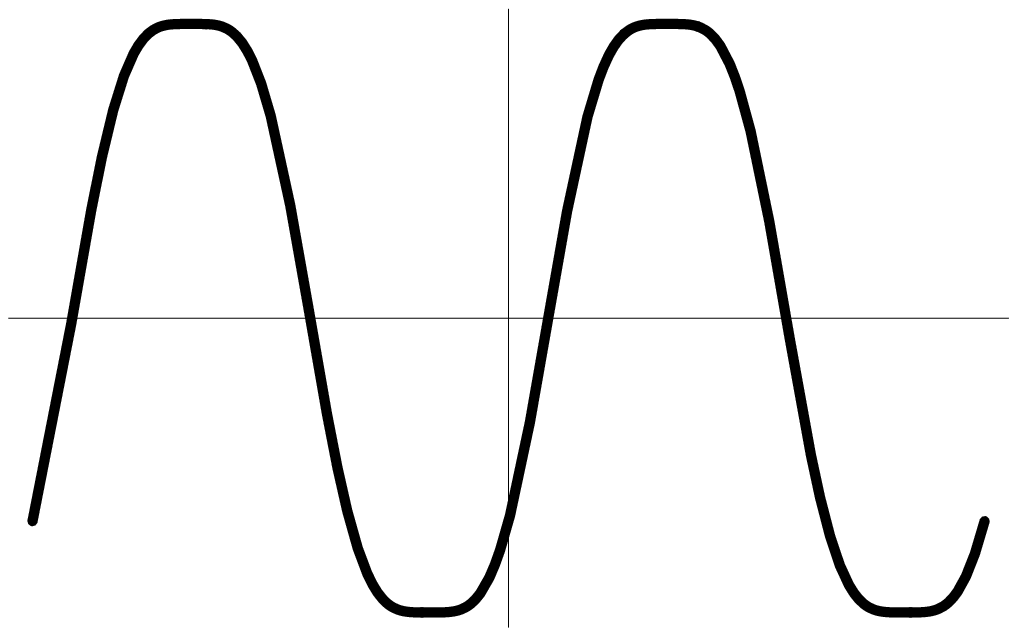,width=\linewidth}
%\caption{(b)}
%\vspace{1mm}
\begin{center}
{\bf (b)}
\end{center}
\end{minipage} \hskip 30pt
\begin{minipage}[b]{.25\linewidth}
\centering\psfig{figure=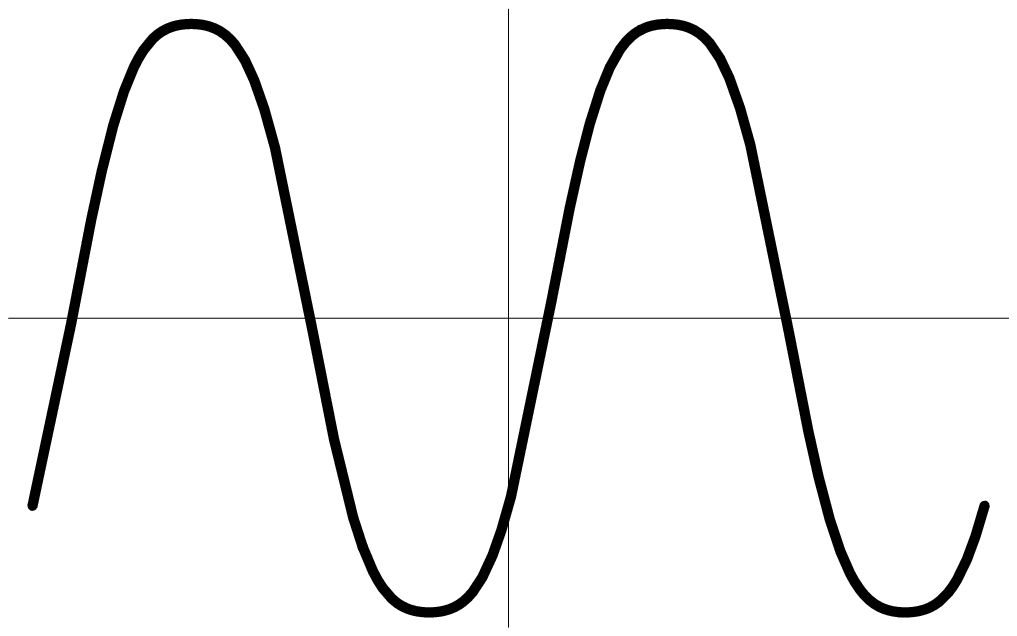,width=\linewidth}
%\caption{(a)}
%\vspace{1mm}
\begin{center}
{\bf (c)}
\end{center}
\end{minipage} 
\caption{{\em Potential of the multiple Sine-Gordon model with $\beta/\alpha =1/3$ and $\delta=\pi/3$
in the vicinity of $\eta_c$: (a) $\eta < \eta_c$, (b) $\eta = \eta_c$ and  (c) $\eta > \eta_c$. }}
\end{figure}

\vspace{3mm}

\resection{Bidimensional Supersymmetric Theories}
 
In this paper we will consider $N=1$ supersymmetric theories. They describe the dynamics of 
a real superfield $\Phi(x,\theta)$ that consists of a scalar field $\varphi(x)$, a real (Majorana) spinor
$\psi_{\alpha}(x)$ and a real auxiliary field $F(x)$, with the expansion 
\EQ
\Phi(x,\theta) \,=\,\varphi(x) + \bar\theta\,\psi(x) + \frac{1}{2} \bar\theta\,\theta F(x) \,\,\,.
\EN 
The space coordinates $x^{\mu}=(x^0,x^1)$ and the the two real Grassmann coordinates 
$\theta_{\alpha} = (\theta_1,\theta_2)$ describe the $N=1$ superspace. We adopt the following 
representation for the $\gamma$-matrices
\EQ
\gamma^0 \,=\,
\left(
\begin{array}{lc}
0 & -i \\
i  & 0 
\end{array}
\right) 
\,\,\,\,\,\,\,\,
,
\,\,\,\,\,\,\,\,
\gamma^1 \,=\,
\left(
\begin{array}{ll}
0 & i \\
i  & 0 
\end{array}
\right)
\EN
The role of $\gamma_5$ is played by 
\EQ 
\gamma_5 = \gamma^0\,\gamma^1 = 
\left(\begin{array}{cc} 1 & 0 \\ 0 & -1 \end{array}\right)
\,\,\,. 
\EN
The charge-conjugation is implemented by the matrix $C_{\alpha \beta}=- (\gamma^0)_{\alpha \beta}$, 
 so that for a charge-conjugate spinor $\psi_{\alpha}^C \equiv C_{\alpha \beta} \,\bar \psi_{\beta} 
 = \psi^{\dagger}_{\alpha}$, with $\bar \psi = \psi^{\dagger} \gamma^0$. Hence, in the above representation, the Majorana spinors $\psi$ and $\theta$ are real. 
 
Under a translation in superspace 
 \EQ
 x^{\mu} \rightarrow x^{\mu} + i \bar\xi \gamma^{\mu} \theta 
\,\,\,\,\,\,\,\,
,
\,\,\,\,\,\,\,\,
\theta_{\alpha} \rightarrow \theta_{\alpha} + \xi_{\alpha} 
\label{translationsusy}
\EN 
the superfield has a variation expressed by 
\EQ
\delta \Phi(x,\theta)\,=\,\bar \xi_{\alpha} \,{\cal Q}_{\alpha}\,\Phi(x,\theta) \,\,\,,
\label{susytran}
\EN 
with ${\cal Q}_{\alpha} = \partial/\partial \bar\theta_{\alpha} + i (\gamma^{\mu} \theta)_{\alpha}\,
\partial_{\mu}$. The most general action invariant under the supersymmetric transformation 
(\ref{susytran}) is given by 
\EQ
{\cal A} \,=\,\int d^2 x\, d^2 \theta \left[
\frac{1}{4} (\bar D_{\alpha} \Phi) \,D_{\alpha} \Phi + W(\Phi) \right] \,\,\,,
\label{susyaction}
\EN 
where $\int d^2\theta \bar \theta \theta = 2$, with the covariant derivative $D_{\alpha}$ 
given by 
\EQ
D_{\alpha} \equiv \frac{\partial}{\partial \bar\theta_{\alpha}} - 
(i \partial_{\mu} \gamma^{\mu} \theta)_ {\alpha} \,\,\,.
\EN 
$W(\Phi)$ is the so-called superpotential, that we assume to be an analytic function of $\Phi$. 
Integrating on the Grassmann variables, one arrives to the following expression of the action 
\EQ
{\cal A}\,=\,\int d^2 x \left\{
\frac{1}{2} \left[ (\partial_{\mu} \varphi)^2 + i \bar\psi \gamma^{\mu} \partial_{\mu}
\psi + F^2 \right] + F\,W^{'}(\varphi) - \frac{1}{2} W^{"}(\varphi) \bar\psi\psi \right\}\,\,\,,
\EN 
where $W^{'}(\varphi) = dW(\varphi)/d\varphi$, etc.  Finally, elimitating the auxiliary field $F$ from 
its algebraic equation of motion, i.e. substituting $F \rightarrow - W^{'}(\varphi)$ in the above 
expression, yields a lagrangian density for a supersymmetric theory given, in its most general form, by 
\EQ
{\cal L} \,=\,
\frac{1}{2} \left[ (\partial_{\mu} \varphi)^2 + i \bar\psi \gamma^{\mu} \partial_{\mu}
\psi \right] -\frac{1}{2} [W^{'}(\varphi)]^2 - \frac{1}{2} W^{"}(\varphi) \bar \psi \psi \,\,\,.
\label{finalsusy}
\EN
Associated to the transformation (\ref{translationsusy}) is the conserved supercurrent 
\EQ
J^{\mu}_{\alpha} \,=\,(\partial_{\nu} \varphi) (\gamma^{\nu} \gamma^{\mu} \psi)_{\alpha} - i F 
(\gamma^{\mu} \psi)_{\alpha} \,\,\,, 
\label{supercurrent}
\EN 
and the conserved supercharges 
\EQ
Q_{\alpha} \,=\,\int dx^1\,J_{\alpha}^0 \,\,\,.
\label{supercharge}
\EN 
Let's also define the stress-energy tensor 
\EQ
{\cal T}^{\mu\nu} \,=\,\frac{i}{2} \bar\psi \,\gamma^{\mu} \partial^{\nu} \psi + 
\partial^{\mu} \varphi \,\partial^{\nu} \varphi - \frac{1}{2} g^{\mu\nu} \left[
(\partial_{\alpha} \varphi)^2 - F^2\right] 
\,\,\,,
\label{stressenergy}
\EN 
and the topological current 
\EQ
\xi^{\mu} \,=\,-\epsilon^{\mu\nu} F\,\partial_{\nu} \varphi \,=\,\epsilon^{\mu\nu} \partial_{\nu} W(\varphi) 
\,\,\,.
\EN 
As shown by Witten and Olive \cite{OliveWitten}, the most general supersymmetry algebra is then
\EQ
\{Q_{\alpha},\bar Q_{\beta}\} \,=\,2 (\gamma_{\lambda})_{\alpha\beta}\,P^{\lambda} + 
2 i (\gamma_5)_{\alpha\beta}\,{\cal Z} \,\,\,,
\label{susyalgebra}
\EN 
where $P^{\lambda} = \int dx^1 {\cal T}^{0 \lambda}$ is the total energy and momentum, 
whereas ${\cal Z}$ is the topological charge 
\EQ
{\cal Z} \,=\,\int dx^1 \xi^0 \,=\,\left[W(\varphi)\right]^{+\infty}_{-\infty} \,\equiv\,
W(\varphi)\mid_{x^1 = +\infty} - W(\varphi)\mid_{x^1 = - \infty} \,\,\,.
\label{topologicalcharge}
\EN 
A convenient and explicit form way of the supersymmetry algebra (\ref{susyalgebra}) is the following 
\EQ
Q_+^2\,=\,P_+ 
\,\,\,\,\,\,\,
,
\,\,\,\,\,\,\,
Q_-^2 \,=\,P_- 
\,\,\,\,\,\,\,
,
\,\,\,\,\,\,\,
Q_+ \, Q_- + Q_- \, Q_+ \,=\, 2 {\cal Z} \,\,\,,
\label{susyconvenient}
\EN 
where we have used the notation $(Q_1,Q_2) \equiv (Q_-,Q_+)$ and $P_{\pm} = P_0 \pm P_1$. 

In the above formulas, ${\cal Z}$ is different from zero only when $W(\Phi)$ admits  several solutions of the equation $W^{'}(\Phi) = 0$, i.e. if the model supports kink excitations that interpolate between different vacua at $x^1 = \pm \infty$. This is also evident from the expression of the topological current $\xi^{\mu}$: this is a purely derivative term which is different from zero only on field configurations that do not vanish at infinity. When it is evaluated on the classical expression of the kinks, ${\cal Z}$ is equal to their mass (see eq.\,(\ref{masskink}) below). Some examples of theories with non-zero ${\cal Z}$ will be considered in Section \ref{susykinksection}.
  
Eq.\,(\ref{finalsusy}) gives the most general lagrangian  of a supersymmetric theory. As evident from its expression, its interactions consists of two terms: the first one, a Yukawa coupling for the fermion ($ V_Y= W^{"}(\varphi) \bar\psi \psi$), the second one, a potential that involves  only the bosonic degree of freedom ($V_B= [W^{'}(\varphi)]^2$).  The last term is intrinsicaly positive and, as well-known \cite{Wittenbreaking}, it can be used to study the spontaneously breaking of supersymmetry: if $[W^{'}(\varphi)]^2$ has zeros, they are the true vacua of the theory and supersymmetry is unbroken; viceversa, if $[W^{'}(\varphi)]^2$ has local minima that are not zeros of this function, supersymmetry is spontaneously broken at these minima. They are called meta-stable vacua
if exists a true vacuum somewhere in the landscape of $V_B(\varphi)$. The possibility of constructing reliable theories with a broken supersymmetry at these meta-stable vacua is a topic  currently  under an active investigation, after the seminal work \cite{Seiberg}. We will further comment on this topic in Section \ref{metastable}. 

\subsection{Superconformal minimal models and deformations}
As for pure bosonic theories, supersymmetric models may present a critical behavior. In the following, our attention will focus on the so-called $N=1$ discrete series ${\cal SM}_m$ \cite{FQSsusy}, identified by combining the conformal invariance of the critical point together with the supersymmetry and the 
unitarity of the models\footnote{In this section we will only remind the properties of these models 
that are useful in the sequel of the paper. For a detailed discussion of the  super-conformal models, in 
addition to \cite{FQSsusy}, see for instance \cite{Qiu,Cappelli,ZamSUSY,zamreview,MSS,Kastor}.}. The central charge of these conformal theories assumes the discrete values 
\EQ
c\,=\,\frac{3}{2} - \frac{12}{m (m+2)} 
\,\,\,\,\,\,\,\,  
,
\,\,\,\,\,\,\,\,  
m=3,4,5,\ldots
\label{central charge}
\EN
The fields ${\cal O}_{r,s;\bar r,\bar s}$, with $1 \leq r, \bar r \leq m-1$ and $1 \leq s, \bar s \leq m+1$, have anomalous dimensions $\eta \,=\,\Delta_{r,s} + \Delta_{\bar r,\bar s}$, where 
\EQ
\Delta_{r,s} \,=\,\Delta_{m-r,m+2-s}\,=\,\frac{[(m+2) r - m s]^2 - 4}{8 m (m+2)} + 
\frac{1}{32} \left[1- (-1)^{r-s}\right] \,\,\,.
\label{conformaldimension}
\EN 
When $r-s$ is an even number, the field ${\cal O}_{r,s;\bar r,\bar s}$ is a primary superfield that belongs to the Neveu-Schwartz representation of the super-algebra. When $r-s$ is instead an odd number, the corresponding field ${\cal O}_{r,s;\bar r,\bar s}$ is an ordinary conformal primary field that creates a representation of the Ramond sector of the super-algebra. 

As shown by Zamolodchikov \cite{ZamSUSY}, the $N=1$ superconformal model ${\cal SM}_m$ (with 
the operator content fixed by the diagonal partition function) can be identified with the Landau-Ginzbrug lagrangian of a real scalar superfield $\Phi$ with the superpotential given, in this case, by $W(\Phi) = \frac{1}{m}\,\Phi^m$.  

The first superconformal model, with $m=3$ and $c=\frac{7}{10}$,  corresponds to the class of universality of the Tricritical Ising Model (TIM) \cite{FQSsusy}. According to the above identification, this model is associated to the superpotential $W(\Phi) =\frac{1}{3}\,\Phi^3$: after eliminating the auxiliary field $F$, it gives rise to the interaction terms 
\EQ
V\, = \, \frac{1}{2} \varphi^4  + \bar \psi \psi \,\varphi \,\,\, .
\label{TIMeffective}
\EN
The next model, with $m=4$ and $c=1$, corresponds to the class of universality of the critical 
Ashkin-Teller model (i.e. a particular gaussian model). Its superpotential $W(\phi) = \frac{1}{4} \Phi^4$ produces the interaction terms 
\EQ
V\,=\, \frac{1}{2}\, \varphi^6 + \frac{3}{2} \bar \psi \psi\, \varphi^2 \,\,\,.
\label{gausseffective}
\EN 
An important difference exists between the superconformal minimal models ${\cal SM}_m$ when 
$m$ is an even or an odd number. To see this, it is useful to computed the Witten index $Tr (-1)^f$ of the superconformal models ($f$ is the fermion number, which assumes the value $f=0$ for bosonic states and $f=1$ for fermionic states). As usual, this can be computed by initially defining the theory on a cylinder \cite{Kastor}: the Hamiltonian on the cylinder is given by $H=Q^2=L_0-c/24$, where $Q$ is the holomorphic compoment of the supercharge, $c/24$ is the Casimir energy on the cylinder and $L_0 = \frac{1}{2\pi i}\oint dz \,z\, T(z)$, with $T(z)$ the holomorphic component of the stress-energy tensor. Since for any conformal state $\mid a\rangle$ with $\Delta > c/24$ there is the companion state $Q\mid a\rangle$ of opposite fermionic parity, their contributions cancel each other in $Tr (-1)^f$ and therefore only the ground states with $\Delta = c/24$ (which are not necessarily paired) enter the final expression of the Witten index. For the minimal models, by using the conformal dimensions given in eq.\,(\ref{conformaldimension}), it is easy to check that there is a non-zero Witten index only for $m$ even: the Ramond field ${\cal O}_{\frac{m}{2},\frac{m}{2} +1}$ of these models has indeed conformal dimension $\Delta = c/24$.  Therefore the lowest superconformal minimal model with a non-zero Witten index is the one with $m=4$, i.e. the gaussian model, whereas the Tricritical Ising Model has a zero Witten index. 

It is well known that the Witten index is related to the supersymmetry breaking problem: if it is nonzero, the supersymmetry cannot be broken and if the supersymmetry is broken it must be zero. This is explicitly confirmed by the examples studied in the literature. In the case of the Tricritical Ising Model, 
perturbing the critical action by the super-operator ${\cal O}_{1,3}$, i.e. inserting the extra term 
$\lambda \,\Phi$ in the supersymmetrix action, produces the bosonic interaction term (see Figure 6)
\EQ
V_B\,=\,\frac{1}{2} (\varphi~^2 + \lambda)^2 \,\,\,.
\label{bosonicTIM}
\EN 
If $\lambda < 0$, the ground state energy is zero and supersymmetry is unbroken: both scalar and fermion fields are massive and the theory has kink excitations \cite{zamkinksusy}. Viceversa, if $\lambda > 0$ the ground state energy is non-zero and supersymmetry is spontaneously broken: the scalar field $\varphi$ becomes massive whereas the fermion stays massless and plays the role of goldstino. In this case, the Landau-~Ginzbrug theory describes the massless flow from the Tricritical Ising model to the critical Ising model \cite{Kastor}. This flow gives rise, in particular, to an integrable field theory with the exact $S$-matrix determined in \cite{ZamTIMflow}. 
Notice that, at the critical point, $V_B$ has a behaviour $V_B \sim \varphi^4$, that obviously does not correspond to the class of universality of the Ising model: in fact, one has to take into account that the superconformal model ${\cal SM}_3$ has also additional interaction relative to its fermionic degree of freedom, i.e. the nteractions of the model are not only expressed by $V_B$. 

\vspace{10mm}
\begin{figure}[h]
\hskip 25pt
\begin{minipage}[b]{.25\linewidth}
\centering\psfig{figure=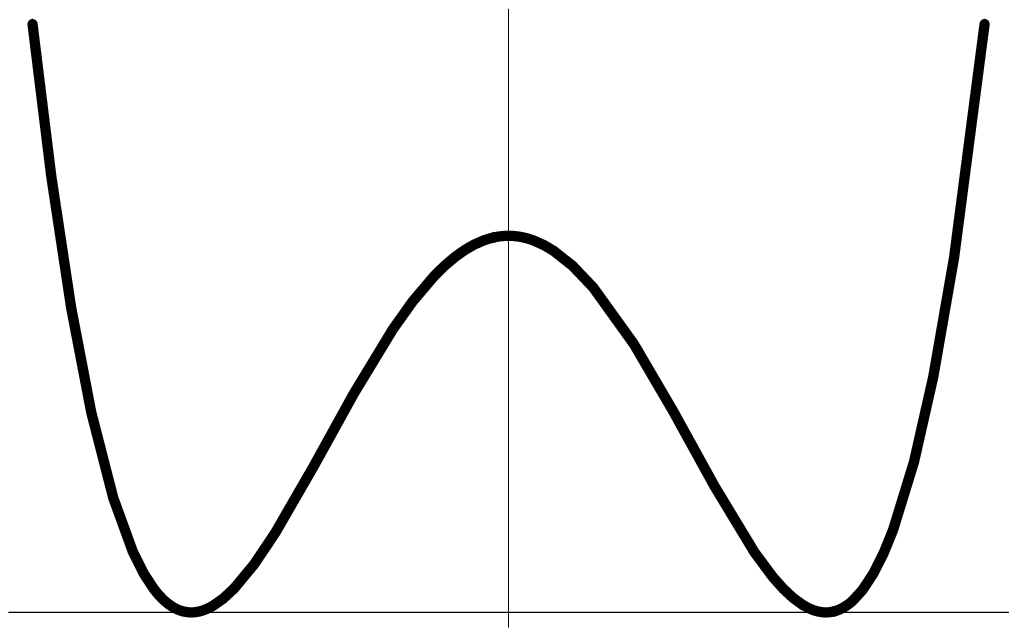,width=\linewidth}
%\caption{(a)}
%\vspace{1mm}
\begin{center}
{\bf (a)}
\end{center}
\end{minipage} \hskip 30pt
\begin{minipage}[b]{.25\linewidth}
\centering\psfig{figure=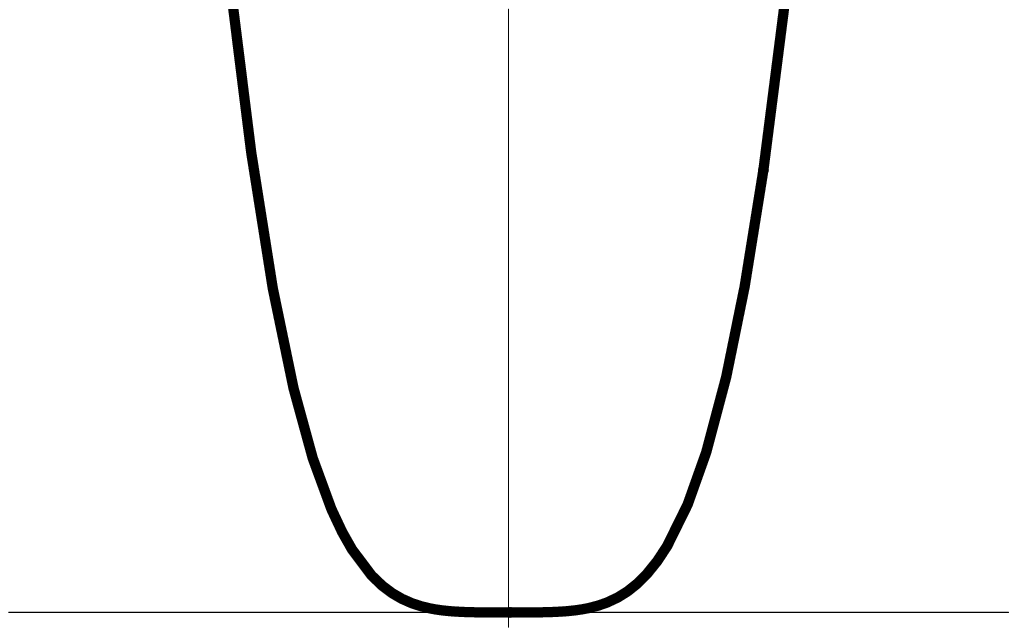,width=\linewidth}
%\caption{(b)}
%\vspace{1mm}
\begin{center}
{\bf (b)}
\end{center}
\end{minipage} \hskip 30pt
\begin{minipage}[b]{.25\linewidth}
\centering\psfig{figure=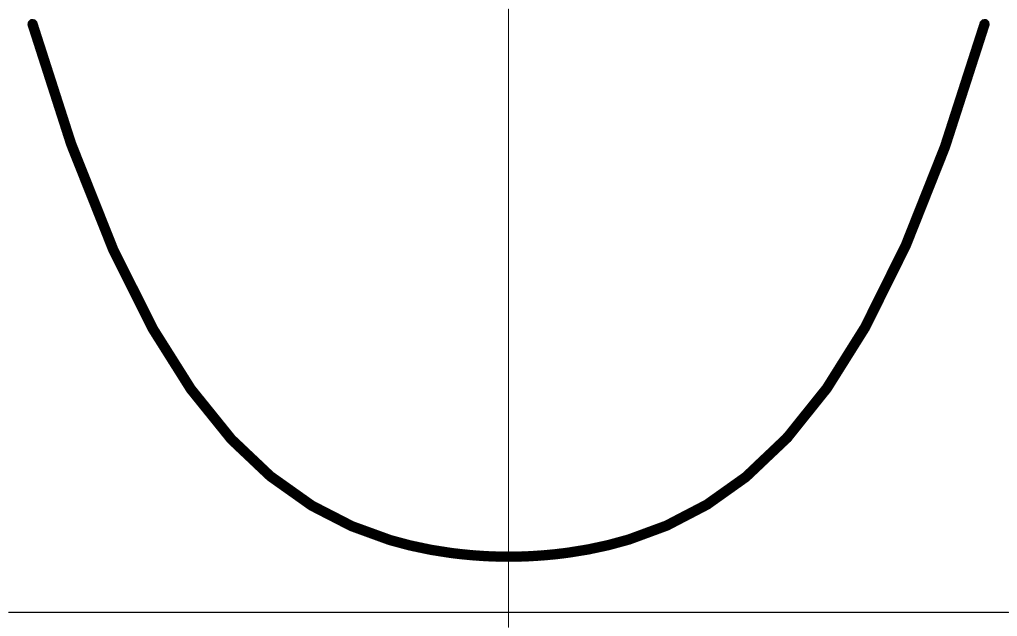,width=\linewidth}
%\caption{(a)}
%\vspace{1mm}
\begin{center}
{\bf (c)}
\end{center}
\end{minipage} 
\caption{{\em Shape of the bosonic interaction term of the Tricritical ising Model by varying $\lambda$: (a) $\lambda < 0$ (susy exact) (b) $\lambda = 0$ (superconformal point)  and  (c) $\lambda > 0$ (broken susy) . }}
\end{figure}
In the gaussian model, the role of the super-operator ${\cal O}_{1,3}$ is played by the term $\Phi^2$. Perturbing the conformal action by means of this operator, one has (see Figure 7)
\EQ
V_B\,=\,\frac{1}{2}\,\varphi^2 \,(\varphi^2 + \lambda )^2 \,\,\,.
\label{gaussianpert}
\EN 
By varying $\lambda$, the landscape of the bosonic potential changes as in Figure 7: in this case, no matter what the sign of $\lambda$ may be, the potential has always at least one zero, i.e. supersymmetry is always unbroken. Notice that at the critical point, $V_B$ has a behaviour at the origin $V_B \sim \varphi^6$. For $\lambda < 0$, the excitations of the model are massive kinks while, for $\lambda > 0$,  they consists of massive scalar and fermion particles localised around the exact vacuum of the theory. 

\vspace{5mm}
\begin{figure}[h]
\hskip 25pt
\begin{minipage}[b]{.25\linewidth}
\centering\psfig{figure=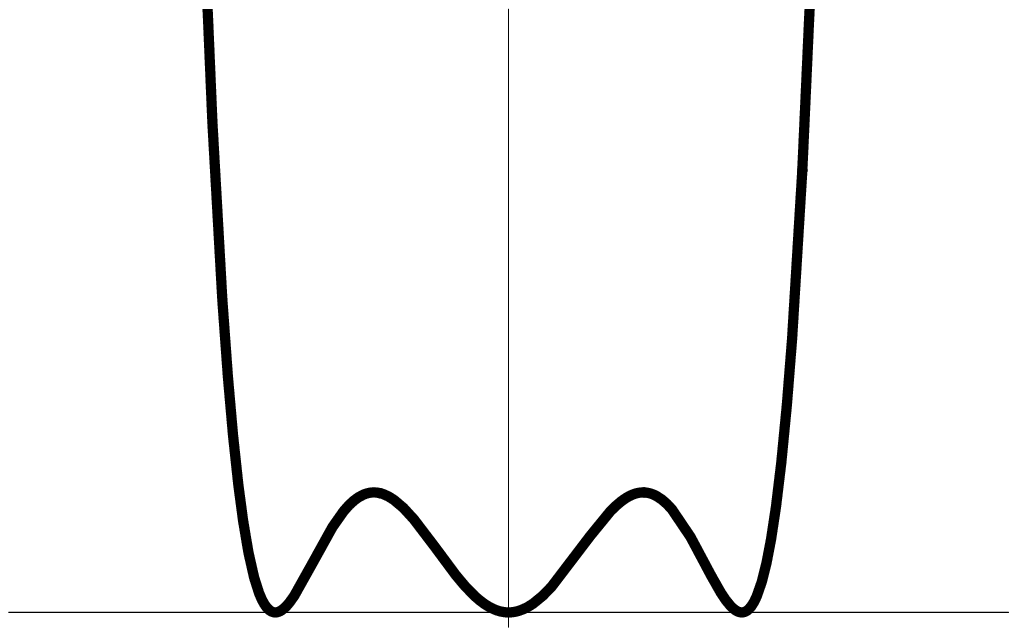,width=\linewidth}
%\caption{(a)}
%\vspace{1mm}
\begin{center}
{\bf (a)}
\end{center}
\end{minipage} \hskip 30pt
\begin{minipage}[b]{.25\linewidth}
\centering\psfig{figure=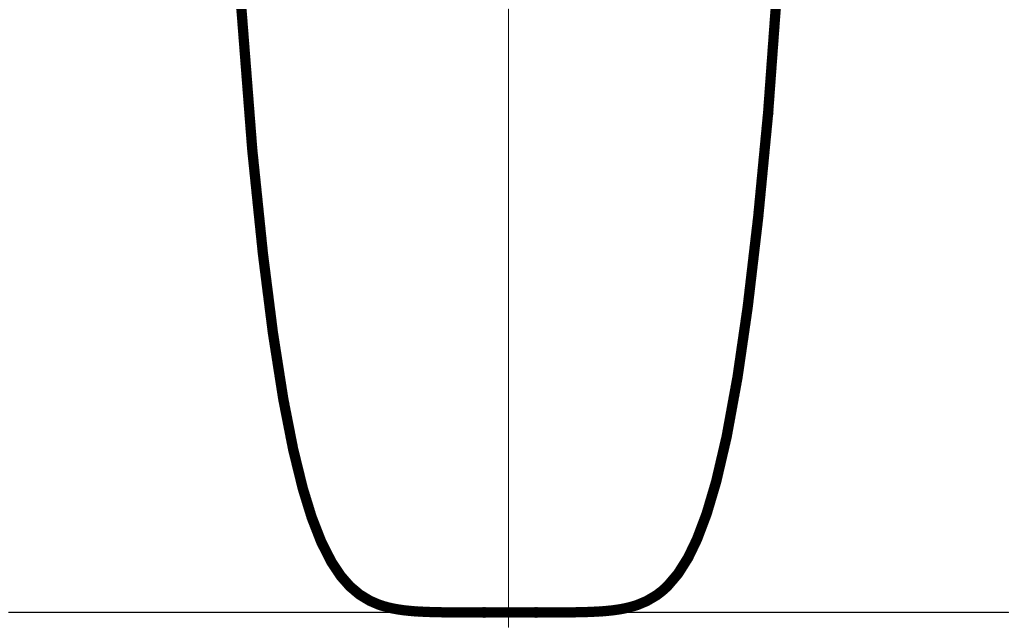,width=\linewidth}
%\caption{(b)}
%\vspace{1mm}
\begin{center}
{\bf (b)}
\end{center}
\end{minipage} \hskip 30pt
\begin{minipage}[b]{.25\linewidth}
\centering\psfig{figure=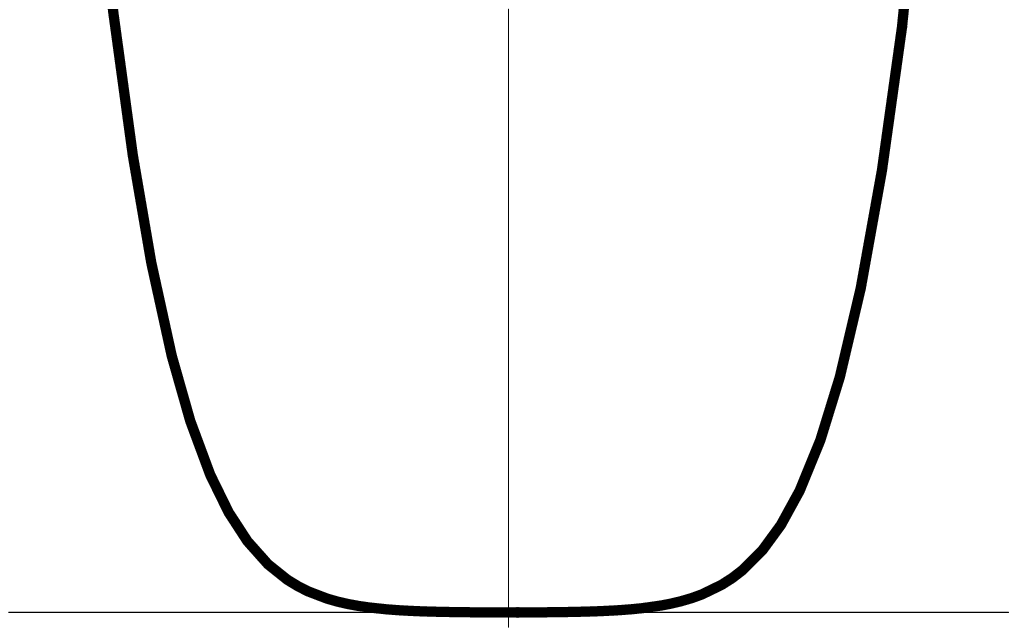,width=\linewidth}
%\caption{(a)}
%\vspace{1mm}
\begin{center}
{\bf (c)}
\end{center}
\end{minipage} 
\caption{{\em Shape of the bosonic interaction term of the gaussian model by varying $\lambda$: (a) $\lambda < 0$ (three degenerate vacua of zero energy: susy exact) (b) $\lambda = 0$ (superconformal point)  and  (c) $\lambda > 0$ (unique vacuum of zero energy:  susy exact) . }}
\end{figure}

\section{Kink in supersymmetric theories} \label{susykinksection}
In the following, as an example of  a supersymmetric integrable theory with kink excitations, we will consider the super Sine-Gordon mode (SSG). This model has been studied by several authors and a great deal is known about it: its quantum integrability, for instance, has been checked in 
\cite{sengupta}, its exact $S$-matrix has been proposed in \cite{AhnSSG} and it follows the general classification given in \cite{Schoutens}, whereas its analysis in the framework of perturbed conformal theories has been carried out in \cite{bdptw}. Its thermodynamics and finite size spectrum has been studied in \cite{tsvelik}. The long-standing and subtle issue of quantum correction to the mass of the solitons has been solved in the series of papers \cite{BPS,shifman}, where it was shown that the soliton states satisfy a BPS property (i.e. their mass being equal to their topological charge) even beyond the tree level.

As action of the SSG we will take 
\EQ
{\cal A} \,=\,\int d^2 x\, d^2 \theta \left[
\frac{1}{4} (\bar D_{\alpha} \Phi) \,D_{\alpha} \Phi -\frac{\mu}{\alpha^2}\, \cos(\alpha \Phi) \right] \,\,\,.
\label{SSGaction}
\EN 
Eliminating the auxiliary field $F$, it gives rise to the Lagrangian 
\EQ
{\cal L}\,=\,
\frac{1}{2} \left[ (\partial_{\mu} \varphi)^2 + i \bar\psi \gamma^{\mu} \partial_{\mu}
\psi \right] -\frac{\mu^2}{2 \alpha^2} \, \sin^2(\alpha\,\varphi) - \frac{\mu}{2}  \cos(\alpha\,\varphi)
 \bar \psi \psi \,\,\,.
\label{finalSSG}
\EN
 In the standard perturbative approach of this model, the parameter $\mu$ (that we assume to be a positive quantity) plays the role of the mass of the scalar excitations nearby each minima of 
$V_B = \frac{\mu^2}{2 \alpha^2} \sin^2(\alpha\varphi)$, that are located at $\varphi_k^{(0)}\,=\,k\,\pi/\alpha$. All these minima have zero energy, i.e.they  are all possible supersymmetric vacua of the quantum theory. A more detailed discussion on the nature of these vacuum states will be done at the end of this section. Notice that the above theory may be also regarded as a massive deformation of the free conformal action with central charge $c = 3/2$, given by a massless bosonic field and a massless fermionic field\footnote{The discussion of this point of view can be found in \cite{bdptw}.}. 

The theory (\ref{finalSSG}) is invariant under the $Z_2$ parity $\varphi \rightarrow - \varphi$ and under the shift 
\EQ
\varphi \rightarrow \varphi + n \frac{2\pi}{\alpha} \,\,\,.
\label{firstperiod}
\EN 
It is also invariant under a half-period shift of the bosonic field
\EQ
\varphi \rightarrow \varphi + n \frac{\pi}{\alpha} \,\,\,, 
\EN  
a pact to change the relative sign of the fermion by $\gamma_5$, $\psi \rightarrow \gamma_5 \,\psi$, i.e. 
\EQ
\psi_1 \rightarrow \psi_1 
\,\,\,\,\,\,
,
\,\,\,\,\,\,
\psi_2 \rightarrow - \psi_2 \,\,\,.
\EN 
In view of the periodicity (\ref{firstperiod}), we can always focus our attention to the interval
 $(0, 2 \pi/\alpha$) of the field $\varphi$. 

The minima of $V_B$ are connected by (supersymmetric) kinks $\mid \,S\,\rangle$, which can be 
identified by the conditions \cite{OliveWitten}
\EQ
(Q_+ \pm Q_-) \,\mid\,S \,\rangle \,=\,0 \,\,\,,
\label{killingBPS}
\EN 
where the $+$ refers to the kink whereas the $-$ to the anti-kink. For the bosonic component of these 
kink states, the above conditions end up in solving the first order differential equations\footnote{See 
\cite{shifman} for an elegant discussion of the classical kink solutions in terms of the superspace formalism.}
\EQ
\frac{d\varphi^{cl}}{dx} \,=\,\pm\,F(\varphi^{cl}) \,=\,\pm\,W^{'}(\varphi^{cl}) \,\,\,.
\label{BPSbound}
\EN 
The kink and the anti-kink solutions (corresponding to $+$ and $-$ respectively) 
of the SSG model are explicitly given by 
\EQ
\varphi^{cl}_{\pm}(x) \,=\,\frac{2}{\alpha} \,\arctan (e^{\pm \mu x}) \,\,\,.
\label{classicalsusykinks}
\EN 
They connect the adjacent vacua $\varphi^{(0)} = 0$ and $\varphi^{(0)} = \pi/\alpha$, all other kinks or antikinks of the model being equivalent to the above. Their classical mass $M$ is expressed in terms of their topological charge, thanks to the identity 
\EQ
P_+  + P_- \,=\, (Q_+ \pm Q_-)^2 \mp 2 {\cal Z} \,\,\,
\label{remarkableidentity}
\EN 
that follows from the supersymmetry algebra (\ref{susyconvenient}): for the kink (anitkink) at rest, the left hand side is given by $P_+ + P_- = 2 M$, whereas the right hand side, using eq.\,(\ref{killingBPS}), is equal to $\mp 2 Z$, therefore \cite{OliveWitten} 
\EQ
M \,=\, |{\cal Z}| \,\,\,.  
\label{masskink}
\EN 
As a matter of fact, the kink and the anti-kink are double degenerate states \cite{jackiwrebbi}. One 
way to see this is to observe that, in the presence of kink solutions, the Dirac equation for the fermion field 
\EQ
i \,\gamma^{\mu}\,\partial_{\mu} \,\psi - \mu \cos(\alpha \varphi^{cl}_{\pm})\,\psi \,=\,0 \,\,\,
\label{dirac}
\EN 
possesses static normalizable solutions $\psi^{(0)}_{\pm}$, localised at the position of the kinks \cite{dhnf,jackiwrebbi,sakagami}.  Namely, they satisfy the static differential equation 
\EQ
i \gamma^1 \,\partial_1 \psi^{(0)}_{\pm} - \mu \cos(\alpha \varphi_{\pm}^{cl})\,\psi^{(0)}_{\pm}\,=\,0 \,\,\,.
\EN 
For the kink backgrounds $\varphi_{\pm}^{cl}$ of the SSG model, their explicit expression is 
given by 
\EQ
\psi^{(0)}_{\pm}(x) \,=\,\frac{N}{\cosh(\mu x)} \,\left(\begin{array}{c} 1 \\ \pm 1 \end{array} \right) \,\,\,,
\label{normalizable}
\EN 
($N$ is a finite normalization constant). These are zero-energy eigenfunctions of the Dirac equation, as it can be seen by looking for solutions of eq.\,(\ref{dirac}) in terms of eigenfunctions, $\psi(x,t) = e^{-i \epsilon_p t}\,\psi_p(x)$. Let's focus the attention, for simplicity, only on the kink configuration, since for the antikink one has an analogous situation. We can expand the fermionic field in terms of its eigenfunctions: taking into account that $\psi$ is a Majorana field, one has  
\EQ
\psi \,=\,a_0\,\psi^{(0)}_+ + \sum_p \left(b_p \,\psi_{p^+} + b^{\dagger}_p \,\psi_{p^-}^c \right)
\,\,\, .
\label{expansion}
\EN 
The operators $a_0$ and $b_p$ satisfy anticommutation relations 
\EQ
\{a_0,a^{\dagger}_0\} \,=\,1 
\,\,\,\,\,\,\,
,  
\,\,\,\,\,\,\,
\{b_p,b^{\dagger}_q\} \, =\,\delta_{p,q} 
\,\,\,,
\EN 
and other anticommutation relation vanish. The operators $b^{\dagger}_p$ and $b_p$ are standard creation and annihilation for the non-zero energy fermions in the presence of the kink. The operator 
$a_0$, associated to the fermion zero mode, has instead a bidimensional representation: operating on the kink state, it produce another kink state with the same energy but different fermion number. Hence, in supersymmetric theories, the kink is a doublet. Denoting these two states by $\mid K \,; \pm \,\rangle$, one has 
\EQ
a_0 \,\mid K \,; \,-\,\rangle \,=\,0 
\,\,\,\,\,\,\,
,
\,\,\,\,\,\,\,
\mid K\,;\,+\,\rangle \,=\, a^{\dagger}_0 \,\mid K\,;\,-\,\rangle \,\,\,.
\label{multiplekink}
\EN 
The same considerations also apply to the antikink state.

In view of the above discussion, the nature of the vacuum states of the SSG model is as follows: every vacuum $\varphi_{2n}^{(0)} \,=\,\frac{2 n \,\pi}{\alpha}$ is non-degenerate, while those located at $\varphi_{2n+1}^{(0)} \,=\,\frac{(2 n +1)\,\pi}{\alpha}$ are double degenerate. As discussed in \cite{bdptw} on the basis of an argument by Zamolodchikov \cite{zamkinksusy}, this can be also argued by the sign of the effective mass term of the fermion field  
\EQ
V_Y \,=\,\frac{\mu}{2} \,\cos(\alpha \varphi) \,\bar\psi\,\psi 
\EN 
at these vacua: when the effective mass is positive, the vacuum is not degenerate, while is double degenerate if the effective mass is negative. Hence, altogether we have the situation shown in 
Figure 8. 

\vspace{10mm}
\begin{figure}[h]
\hskip 40pt
\begin{minipage}[b]{.35\linewidth}
\vspace{-5mm}
\centering\psfig{figure=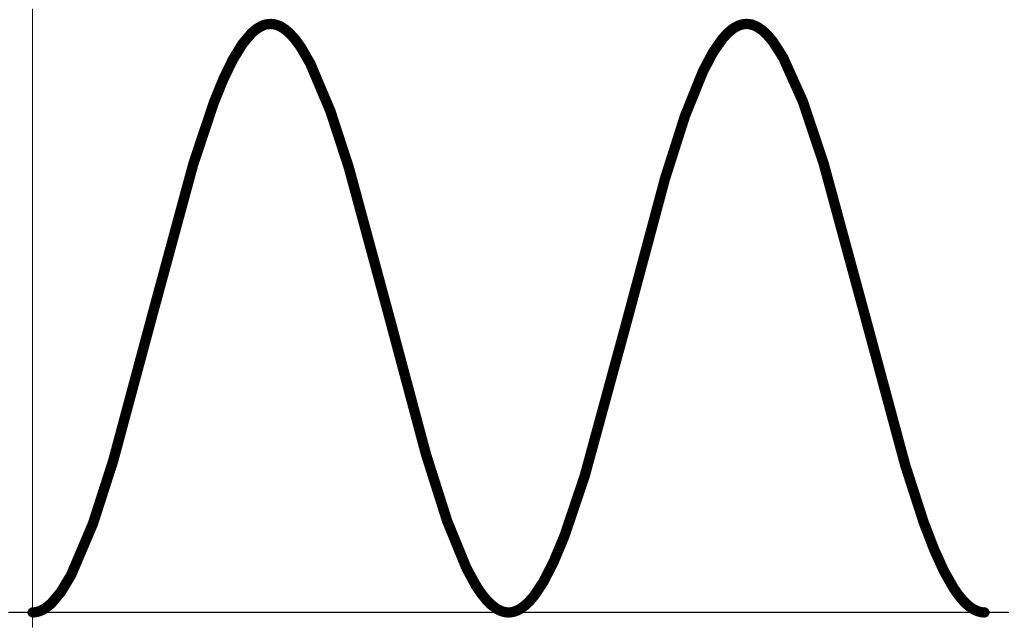,width=\linewidth}
%\caption{(a)}
%\vspace{1mm}
\begin{center}
{\bf (a)}
\end{center}
\end{minipage} \hskip 45pt 
\begin{minipage}[b]{.40\linewidth}
\vspace{-3mm}
\centering\psfig{figure=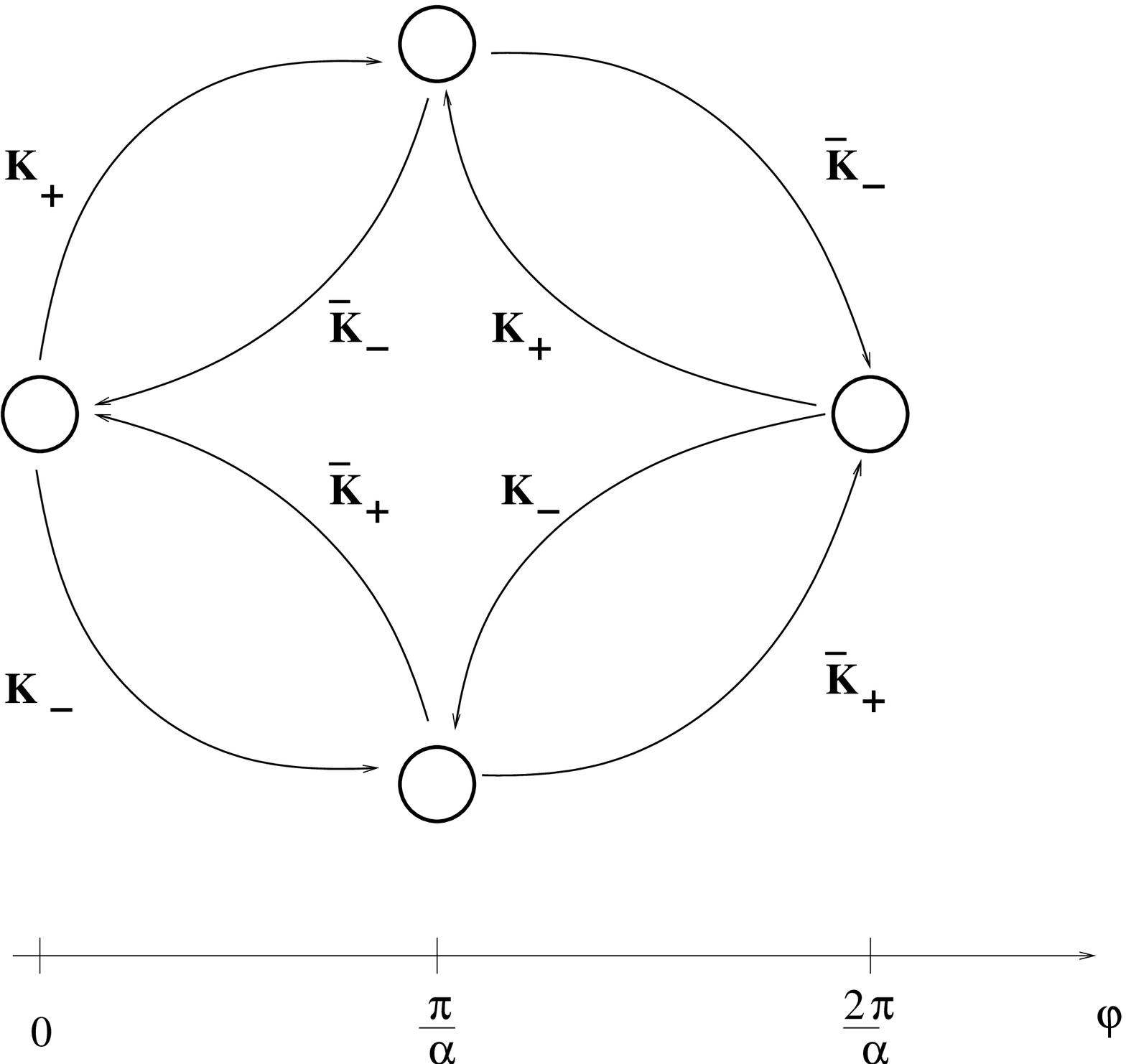,width=\linewidth}
%\caption{(b)}
%\vspace{1mm}
\begin{center}
{\bf (b)}
\end{center}
\end{minipage}
\caption{{\em (a) Plot of $V_B$ of the SSG model in the range $(0,2\pi/\alpha)$. (b) Structure of the vacua and of the kink excitations.}}
%\vspace{5mm}
%\begin{center}
%{\bf Figure 2} 
%\end{center}
\end{figure}

\resection{Multi-frequency Super Sine-Gordon Model} 

Let's now deform the supersymmetric action (\ref{SSGaction}) by inserting another trigonometric 
term, i.e let's consider the action 
\EQ
{\cal A} \,=\,
\int d^2 x\, d^2 \theta \left[
\frac{1}{4} (\bar D_{\alpha} \Phi) \,D_{\alpha} \Phi -\frac{\mu}{\alpha}\, \cos(\alpha \Phi) 
- \frac{\lambda}{\beta} \,\cos(\beta \Phi)
\right] \,\,\,,
\label{SSGdeformed}
\EN 
where we have opportunately rescaled the couplings. By using the super-Coulomb gas formalism 
\cite{coulomb} and the operator product expansion of the (super) exponential operators present in 
(\ref{SSGdeformed}), one can establish the renormalizability (i.e. the stability) of this action in the range of the parameters
\EQ
\alpha^2 < 8 \pi 
\,\,\,\,\,\,\,
,
\,\,\,\,\,\,\,
\beta^2 < 8 \pi 
\,\,\,\,\,\,\,
,
\,\,\,\,\,\,\,
\alpha \,\beta < 4 \pi \,\,\,.
\EN 
In the following we assume the validity of these inequalities. Moreover, we restrict the attention 
to the case in which the ratio $\omega$ of the two frequencies is a rational number 
\EQ
\omega\,\equiv \,\frac{\beta}{\alpha} \,=\,\frac{p}{q} \,\,\,, 
\label{ratiofrequencies}
\EN 
with $p$ and $q$ two co-prime natural numbers, $p< q$. Since any  irrational number can be approximated with arbitrary precision by the sequence of rational approximants given by its continued fraction expressions, the case when $\omega$ is irrational may be regarded as a particular limit of the rational situation.

Since the two trigonometric interactions enter the action (\ref{SSGdeformed}) in a symmetric way, 
there are two natural perturbative regimes of the theory\footnote{As in the bosonic case \cite{DM}, 
the quantum theory actually depends on the appropriate dimensionless combination of the two couplings.}: $\lambda \rightarrow 0$ (with $\mu$ fixed) or $\mu \rightarrow 0$ (with $\lambda$ fixed). Below we will deal with the first regime, since the other can be  simply recovered by swapping the role of the two operators. 

The question we would like to address concerns the fate of the kinks of the original SSG: are they going to be confined once the new interactions is switched on? Or, on the contrary, are they going to remain stable excitations of the perturbed action (\ref{SSGdeformed})? To answer these questions, let us first of all eliminate the auxiliary field $F$ and then focus the attention on the resulting bosonic interaction $V_B$ associated to the action (\ref{SSGdeformed})
\EQ
V_B(\varphi,\lambda) \,=\,\frac{1}{2} \left[\mu \,\sin(\alpha \varphi) + \lambda \, \sin(\beta \varphi)\right]^2
\,\,\,.
\label{pot}
\EN 
To simplify the following formulas, let us rescale the field $\varphi \rightarrow \varphi/\alpha$ and 
the coupling constant $\lambda \rightarrow \lambda/\mu$ in such a way that, up to an overall constant,  
 $V_B$ assumes the form 
 \EQ
\hat V_B(\varphi,\lambda) \,=\,\left[\sin \varphi + \lambda \,\sin \left(\omega\,\varphi \right)\right]^2\,=\,
\left[\sin \varphi + \lambda \,\sin \left(\frac{p}{q}\,\varphi \right)\right]^2 \,\,\,.  
\label{newpot}
\EN 
The first observation is that, no matter what are the values assumed by  $\lambda$, the origin $\varphi^{(0)} = 0$ is always a zero of this expression, i.e. this model always possesses a true supersymmetric vacuum. We will comment later on the possibility of having meta-stable vacua, i.e. vacua where the supersymmetry is spontaneously broken, by varying $\lambda$. 

The second observation is that, at lowest order in $\lambda$, all the vacua of the original SSG potential continue to remain degenerate, i.e. the original kinks are always stable at weak coupling! The simplest way to show this, is to follow the evolution of the zeros of (\ref{newpot}) when we switch on $\lambda$. At $\lambda =0$, they are placed at $\varphi^{(0)}_n = n \pi $.  Switching on $\lambda$, we can look for the new location of the zeros in power-series in $\lambda$, i.e. 
\EQ
\varphi^{(0)}_n \simeq n \pi + \lambda \epsilon^{(1)}_n  + \cdots 
\label{shiftzeros}
\EN 
Substituting this expression into the equation $\hat V_B(\varphi,\lambda) = 0$, it is easy to prove that, at the first order in $\lambda$, $V_B(\varphi,\lambda)$ has the same zeros of $V_B(\varphi,0)$. This can be seen as a simple consequence of the fact that $V_B$ is given, in supersymmetric theory, by the square of a function: namely, if $V_B(x) = [f(x)]^2$ and $f(x)$ has zeros at $x=x_1,x_2,\ldots$ in the interval $I$, perturbing this function by $\lambda g(x)$ one has $V_B(x,\lambda) = [f(x) + \lambda g(x)]^2 \simeq [f(x)]^2 + 2 \lambda f(x) g(x)$. Therefore, if $g(x)$ does not have zeros in the interval $I$, at the lowest order in $\lambda$, $V_B(x,\lambda)$ has the same zeros of $V_B(x)$ in $I$. The actual shift of the zeros can be computed by imposing that $V_B(\varphi,\lambda)$ has zeros also at the second order in $\lambda$: the consistency of this request comes from the result there is always a solution in terms of $\epsilon^{(1)}_n$, given by
\EQ
\epsilon^{(1)}_n = (-1)^n \sin\frac{n p \pi}{q} \,\,\,.
\label{actualshift}
\EN 
Hence, at weak coupling, the zeros $\varphi^{(0)}_n$ shift their position by $\epsilon^{(1)}_n$ but they do not disappear.     

The stability of the kinks at the lowest order in $\lambda$ and the role played by supersymmetry becomes even more evident if we apply the considerations of  Section \ref{residueconf}. For doing so, 
 the first thing is to identify the operator that, at lowest order in $\lambda$, spoils the integrability of the original model. It is easy to see that the role of $\Upsilon$ is played in this case by the operator  
\EQ
\Theta \,=\, \sin\varphi \,\sin(\omega \varphi) + \omega \bar \psi \psi \cos(\omega \varphi) 
\,\,\,.
\label{perturbingop}
\EN 
To compute its semi-local index $\gamma_{\Theta,a}$ on the kink state, we can use the fact 
that the fermionic and the bosonic field are components of the same superfield and that 
the exponential operators $e^{i \eta \varphi}$ has a semi-local index $\gamma = \frac{\eta}{2}$ 
with respect to the kink of this theory\footnote{Notice that this is half the semi-local index of the bosonic Sine-Gordon case,  since the supersymmetric bosonic potential $\sin^2\varphi$ has a frequency that is twice the frequency of the usual Sine-Gordon model.}. If we quantize the theory by choosing as  vacuum 
the zero at the origin, for the residue of the Form Factor of  $\Theta$ that controls the confinement of the kinks we have 
\EQ
{\mbox Res}_{\theta = \pm i \pi} \,F^{\Theta}_{a\bar a}(\theta) \,=\,
\left[1+ \cos (\pi \omega)\right] \,
\,_0\langle 0|\Theta(0)|0\rangle_0 \,\,\,.
\label{residuesusy}
\EN 
Notice that the above quantity consistently vanishes if $\omega =1$, i.e. if we had decided to perturb the action by the same trigonometric term that enters its original definition: in this case the deformation just redefines the original coupling constant $\mu$ and this cannot have consequences on the confinement of the kink. For all other values of $\omega$, the term $[1 + \cos(\pi \omega)]$ is different from zero 
but the residue is nevertheless zero, because it is the vacuum expectation value of $\Theta$  that vanishes in this case! This is a consequence of the unbroken supersymmetry of the theory: in fact, 
 $\Theta$ is nothing else that an additional term in the trace of the stress-energy tensor of theory and, 
 since the supersymmetry is exact, from $Q_{\pm} \mid 0\,\rangle_0 \,=\, 0$ simply follows 
 $_0\langle 0 | \Theta(0) | 0 \rangle_0 \,=\,0$ \cite{Freund}. It is interesting to notice that the absence of the confinement of the kinks is also independent from the rational or irrational nature of $\omega$. 
 
 From a naive point of view, the vanishing of the vacuum expectation value of the perturbing operator 
 $\Theta$  is due to the simultaneous presence of a bosonic and a fermion term in its expression: in this case, the vacuum expectation value of the fermionic term tends to cancel the bosonic one. The exact  cancellation obviously happens for an opportune fine-tuning of the relative couplings of these terms -- a fact that simply expresses the supersymmetry invariance of the theory. In the usual multi-frequency Sine-Gordon, this cancellation is absent for the only presence of the bosonic term: the result is then, in general, the confinement of its kinks. 

The vanishing of the residue (\ref{residuesusy}) implies the absence of confinement of the kinks at weak coupling but, obviously, this does not imply that their mass remain untouched. As a matter of fact, the masses change and, in particular, the kinks are no longer degenerate when $\lambda$ is switched on. Namely, the original kink excitations split into families of  {\em long-kinks} and {\em short kinks}: the former overpass higher barriers, the latter lower barriers (see, for instance, Figure 9).  The actual computation of the new masses of the kinks can be done by employing their topological charges, i.e. by using eqs.\,(\ref{topologicalcharge}) and (\ref{masskink}), together with the shift of the zeros given by (\ref{actualshift}).

\resection{Phase transition and meta-stable states}\label{metastable}
The analysis done in the previous section applies in the weak coupling regime $\lambda \rightarrow 0$. One may wonder, however, what is the evolution of the spectrum\footnote{The SSG model has, in addition to the kinks, also bound states thereof. In this paper, for the sake of simplicity, we focus our  attention only on the kink excitations of the theory.} of the theory by varying $\lambda$ to finite values or even pushing it  to the strong coupling limit  $\lambda \rightarrow \infty$. The result , as we will see, is quite interesting. 

Consider the bosonic and the fermionic potential terms of the theory, in their rescaled form 
\begin{eqnarray}
\hat V_B & \,=\,& \left[\sin \varphi + \lambda \,\sin \left(\frac{p}{q}\,\varphi \right)\right]^2 \,\,\,.
\label{newper} \\
\hat V_Y & \,=\, &  \left[\cos\varphi + \lambda \,\frac{q}{p} \,\cos\left(\frac{p}{q}\,\varphi
\right) \right] \, \bar \psi \psi\,\,\,.
\nonumber 
\end{eqnarray}
In the presence of $\lambda$, the periodicity of the theory is no longer equal to $2 \pi$ but has become $2 \pi q$.  Therefore it is convenient to plot $V_B$ on this extended interval, as in Figure 9.
 
 \vspace{3mm}
\begin{figure}[h]\label{SGfigure2}
\hskip 5pt
\begin{minipage}[b]{.45\linewidth}
%\centering\psfig{figure=m1.eps,width=\linewidth}
\centering\psfig{figure=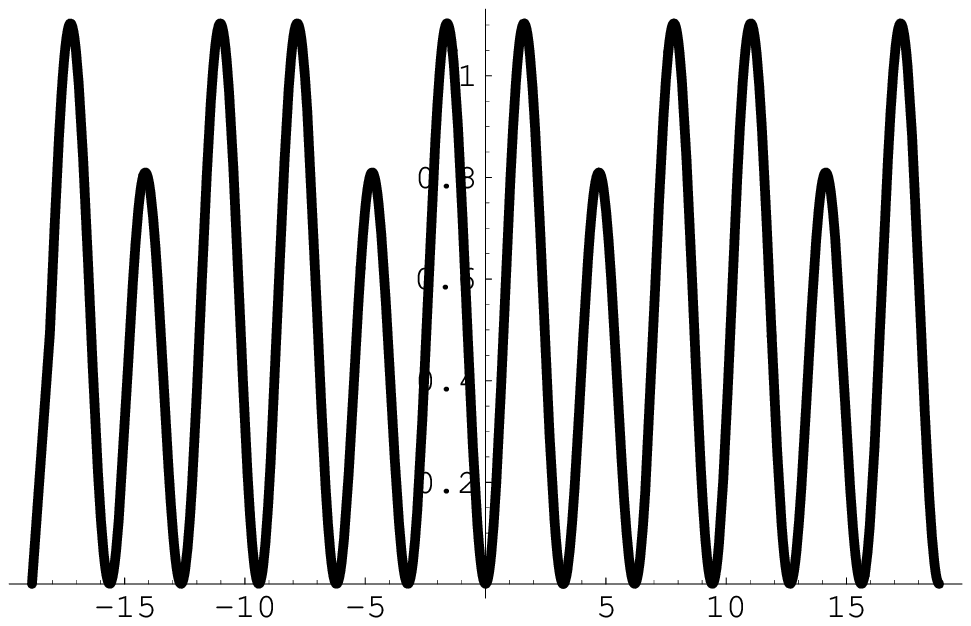,width=\linewidth}
%\caption{(a)}
%\vspace{1mm}
\begin{center}
{\bf (a)}
\end{center}
\end{minipage} \hskip 30pt
\begin{minipage}[b]{.45\linewidth}
\centering\psfig{figure=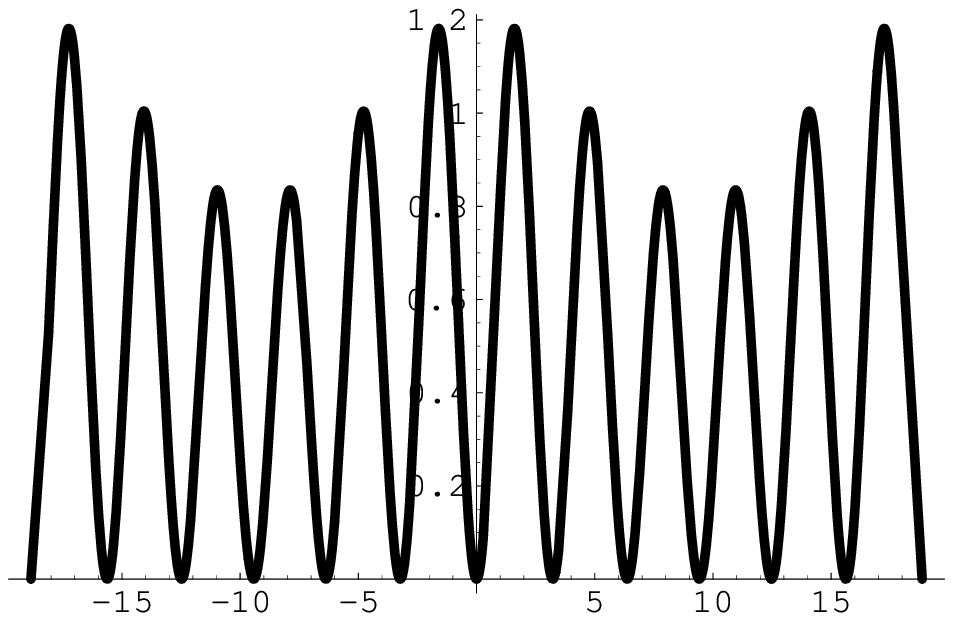,width=\linewidth}
%\centering\psfig{figure=m2.eps,width=\linewidth}
%\caption{(b)}
%\vspace{1mm}
\begin{center}
{\bf (b)}
\end{center}
\end{minipage}
\caption{{\em Plot of $V_B$ with $\lambda =0.1$ in the interval $(-2 \pi q, 2\pi q)$ for $q=3$:  (a) $p=1$ and (b) $p=2$.}}
%\vspace{5mm}
%\begin{center}
%{\bf Figure 2} 
%\end{center}
\end{figure}
The evolution of the spectrum by varying $\lambda$ is constrained by a simple argument. At $\lambda = 0$, one has the sequence of the zeros of the first trigonometric term, which become slightly displaced from their original position when $\lambda$ is small, without changing though their number.  Denoting by $N_z(\lambda)$ the number of zero at a given $\lambda$ in the enlarged interval $(0, 2\pi q)$, we have 
\EQ 
N_z(0) \,= \, 2 q +1\,\,\,
\EN 
(in the above number we have also included the zero at the origin). On the other hand, when $\lambda \rightarrow \infty$, the number of zeros becomes 
\EQ
\lim_{\lambda \rightarrow \infty} N_z(\lambda)\, =\,  2 p +1\,\,\,,
\EN 
i.e. the number of zeros in the interval $(0,2 \pi q$) of the second trigonometric term in $V_B$.  Hence, by varying $\lambda$, there should be a variation of the number of zeros equal to 
 \EQ
 \Delta N_z\, =\, 2 (q-p)\,\,\,.
 \EN 
 Since the kinks own their existence to the zeros,  a variation of their number implies that certain kinks 
 will disappear by moving $\lambda$ from $\lambda = 0$ to $\lambda =\infty$. Since they are topological excitations, their disappearance signals that certain phase transitions will take place in the model at some critical values of the coupling, $\lambda = \lambda_c^{(n)}$. At these critical values, the system has massless excitations that will rule its long-distance behaviour.  As discussed below, $\lambda_c^{(n)}$  are identified as the values where $N_z(\lambda)$ jumps by a step of 2 units (Figure 10). 
 
% \vspace{5mm}

\begin{figure}[h]
\hspace{45mm}
\vspace{10mm}
\psfig{figure=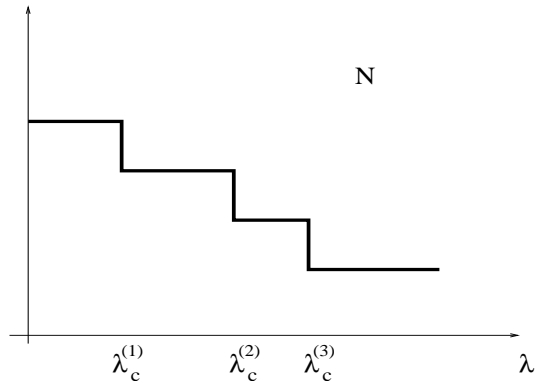,height=5cm,width=7cm}
\vspace{1mm}
\caption{{\em $N_z(\lambda)$ vs $\lambda$, with the sequence of the critical values $\lambda_c^{(n)}$}}
\label{jumps}
\end{figure}

The way in which the number of zeros changes, by varying with continuity $\lambda$, is that pairs of zeros collide and then move on immaginary values. When this happens, the barrier between the two 
colliding vacua vanishes and, correspondingly, the kinks that connect them become massless.  
Notice that, in the evolution of the zeros, the ones placed at  $\varphi^{(0)} = 0$, $\varphi^{(0)} = \pi q$ and $\varphi^{(0)} = 2 \pi q$ will never disapper.  

There is, however, a different way in which the disappearance of the zeros is implemented, according whether  $(p-q)$ is an even or an odd number. In fact, when $(q-p)$ is an odd number, among the pairs of colliding zeros there will always be a couple of them placed just on the right and on the left position of $\varphi^{(0)} = \pi q$:  at some critical value $\hat\lambda_c$, these zeros will collide but strangling, in the meantime, the zero at $\pi q$ that is in between. When $(q-p)$ is an even number, this will not happen. Hence,  the conclusion is  the following\footnote{The analysis refers to $\lambda > 0$. If $\lambda$ is negative there is a swap of the two cases.}: 
\begin{itemize}
\item When $(q-p)$ is an even number, there will be a sequence of phase transitions that will recall the phase transition of the Tricritical Ising Model, i.e. a phase transition with a local spontaneously supersymmetry breaking (see eq.\,(\ref{bosonicTIM}) and the relative discussion). At the critical points, the local form of the potential $V_B$ at the position $\tilde\varphi$ of the colliding zeros is given by $V_B \sim (\varphi - \tilde\varphi)^4$.
\item   
 When $(q-p)$ is  an odd number, in addition to a sequence of TIM-like phase transitions, at a certain value of the coupling, $\lambda = \hat\lambda_c$, there is also a phase transition localized at the vacuum $\varphi = \pi q$. This phase transition will recall the phase transition of the gaussian model, i.e. a phase transition without a local spontaneously supersymmetry breaking (see eq.\,(\ref{gaussianpert}) and the relative discussion). At  $\lambda = \hat \lambda_c$ the local form of the potential $V_B$ at $\varphi = \pi q$ is given by   $V_B \sim (\varphi - \pi q)^6$. 
 \end{itemize} 
 
 \noindent
 In order to make more transparent the above analysis, it is useful to consider explicitly two examples:   
 the first refers to $\omega = \frac{1}{3}$, i.e. $(q-p) =2$, the second instead 
 to $\omega = \frac{2}{3}$, with $(q-p) = 1$.

\vspace{5mm}
\begin{figure}[h]
\hskip 25pt
\begin{minipage}[b]{.25\linewidth}
\centering\psfig{figure=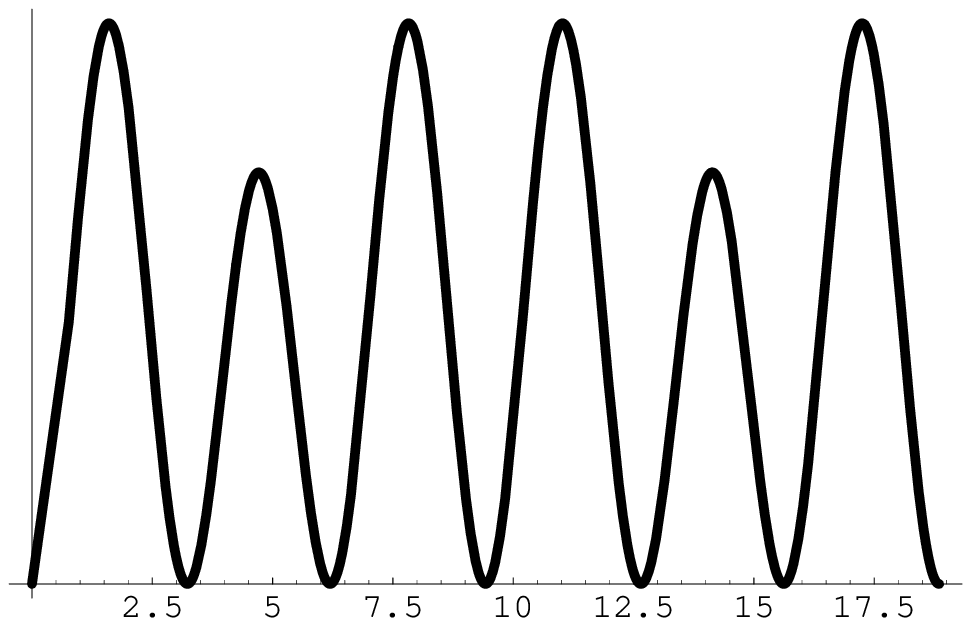,width=\linewidth}
%\caption{(a)}
%\vspace{1mm}
\begin{center}
{\bf (a)}
\end{center}
\end{minipage} \hskip 30pt
\begin{minipage}[b]{.25\linewidth}
\centering\psfig{figure=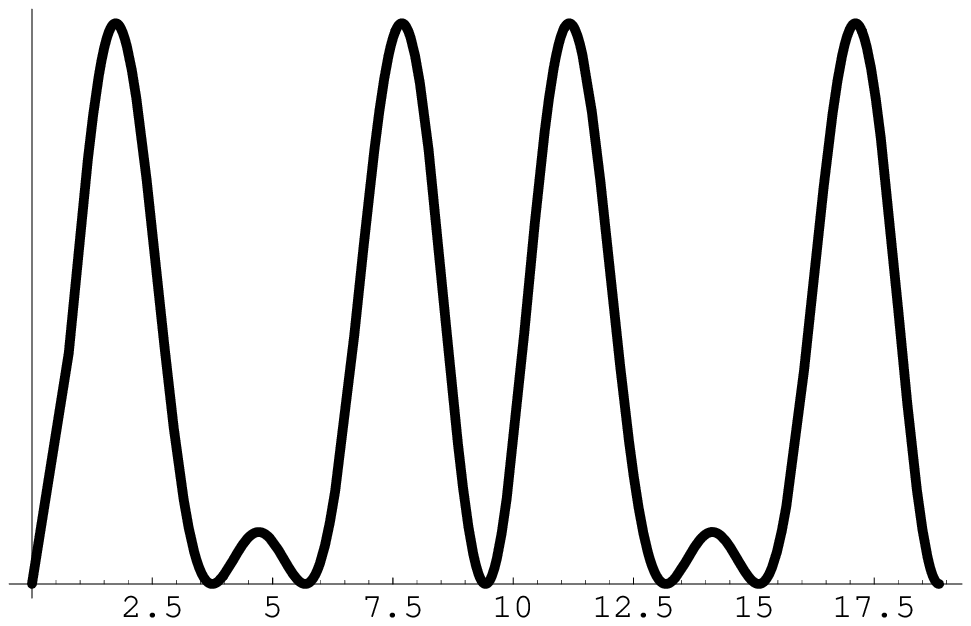,width=\linewidth}
%\caption{(b)}
%\vspace{1mm}
\begin{center}
{\bf (b)}
\end{center}
\end{minipage} \hskip 30pt
\begin{minipage}[b]{.25\linewidth}
\centering\psfig{figure=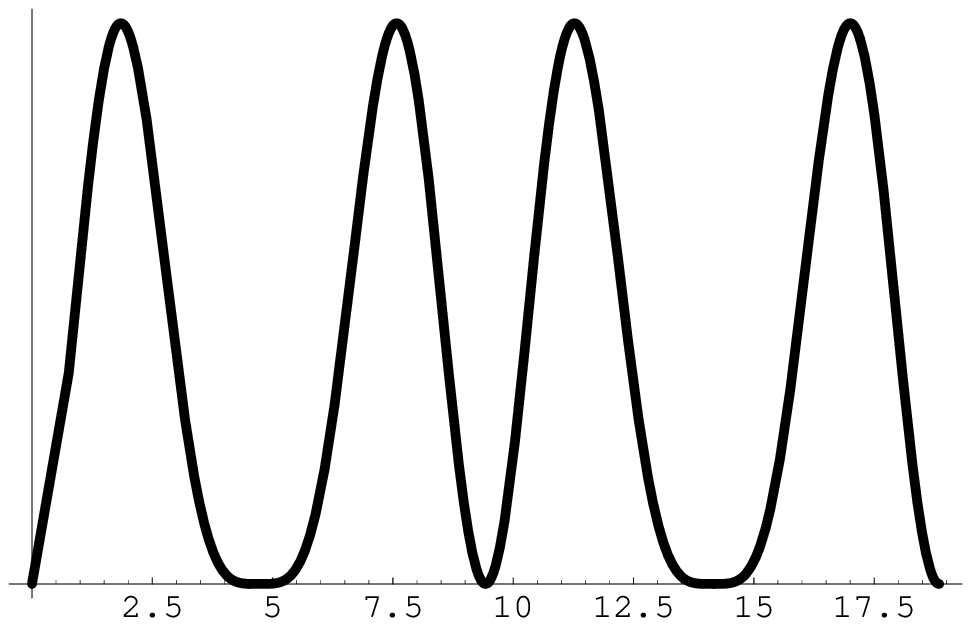,width=\linewidth}
%\caption{(a)}
%\vspace{1mm}
\begin{center}
{\bf (c)}
\end{center}
\end{minipage} \hskip 30pt
\end{figure}
\vspace{3mm}
\begin{figure}[h]
\hskip 25pt
\begin{minipage}[b]{.25\linewidth}
\centering\psfig{figure=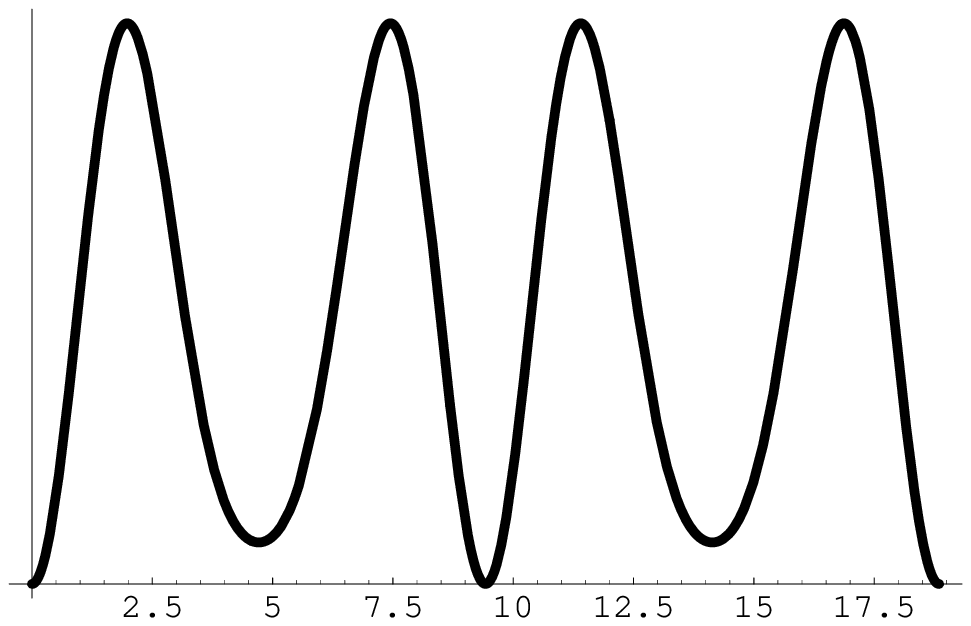,width=\linewidth}
\begin{center}
{\bf (d)}
\end{center}
\end{minipage} \hskip 30pt
\begin{minipage}[b]{.25\linewidth}
\centering\psfig{figure=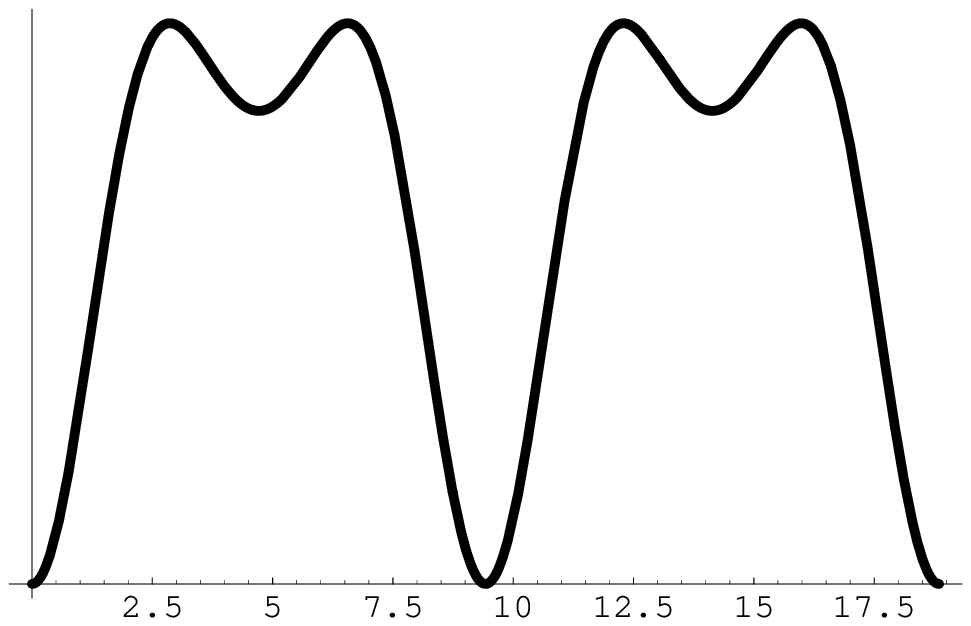,width=\linewidth}
%\caption{(b)}
%\vspace{1mm}
\begin{center}
{\bf (e)}
\end{center}
\end{minipage} \hskip 30pt
\begin{minipage}[b]{.25\linewidth}
\centering\psfig{figure=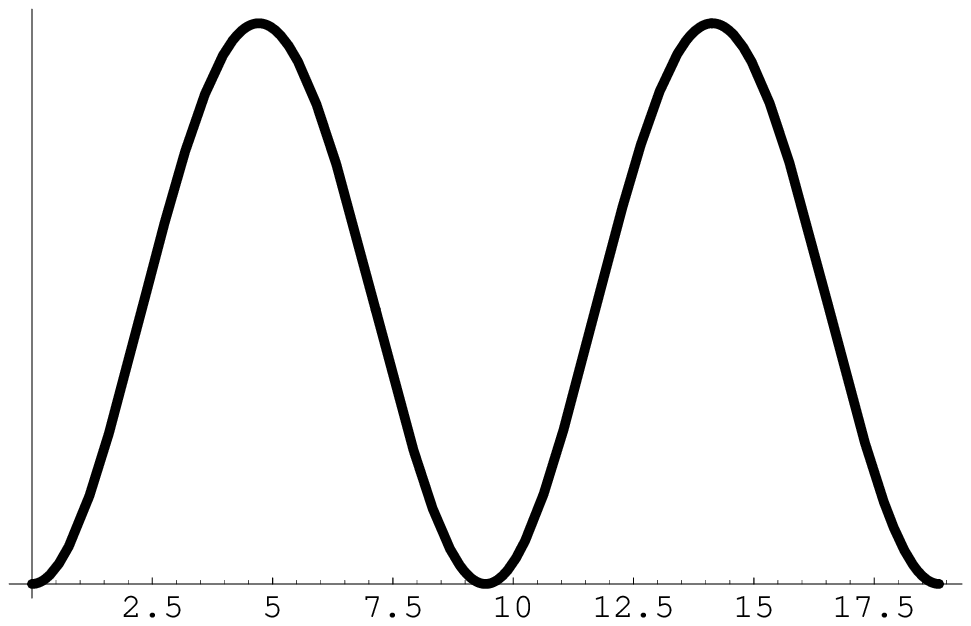,width=\linewidth}
%\caption{(a)}
%\vspace{1mm}
\begin{center}
{\bf (f)}
\end{center}
\end{minipage} 
\caption{{\em Plot of $V_B$ in the interval $(0, 6 \pi)$ for $\omega = 1/3$ by varying  $\lambda$: (a) small value of $\lambda $; (b) $\lambda < \lambda_c$; (c) $\lambda = \lambda_c$; (d) $\lambda > \lambda_c$; (e) $\lambda \gg \lambda_c$; (f) $\lambda \rightarrow \infty$ . }}
\end{figure}

  Let us consider first the case $\omega = \frac{1}{3}$. By varying $\lambda$, the landscape of $\hat V_B$ changes as shown in the sequence of the pictures of Figure 11. Observe that, just before reaching $\lambda_c$, there are two zeros  in the interval $(0,3 \pi)$  that are going to merge together (obviously there are similar zeros in all other intervals obtained by periodicity). Let's call these vacua $\varphi_1^{(0)}$ and $\varphi_2^{(0)}$. At $\lambda = \lambda_c$ the barrier between these two zeros vanishes and, soon after $\lambda_c$, a meta-stable vacuum is created at their middle position $\varphi = \varphi_{mv}$. For $\lambda$ close to $\lambda_c$, the shape of $V_B$ at $\varphi_{mv}$ can be parameterized similarly to the bosonic potential of the Tricritical Ising Model 
  \EQ
  V_B \,\sim \,\left[(\varphi - \varphi_{mv})^2 + (\lambda - \lambda_c) \right]^2 \,\,\,. 
  \EN
Since at $\lambda = \lambda_c$ some topological excitations disappear, a phase transition takes place in the model: if we had originally decided to quantize the theory either at $\varphi_1^{(0)}$ or at $\varphi_2^{(0)}$, the effective dynamics  seen by any of these two vacua appears as a cascade of massless flows: the first, at $\lambda = \lambda_c$, from $c = 3/2$ to $c=7/10$ (the central charge of TIM) and soon after $\lambda_c$, from $c=7/10$ to $c=1/2$, the last value being the central charge of the Ising model. This jumps in the central charge are consistent with the $c$-theorem \cite{cth} and with the unitarity of the theory. Although the zero at the origin always leaves the supersymmetry of the model exact, the quantization around the meta-stable vacuum state at $\varphi_{mv}$ realises however a spontaneously  supersymmetric breaking situation, with a spectrum given by a massive scalar and a massless fermion. This can be explicitly checked by studying $V_Y$: for $\lambda > \lambda_c$ (and for all the range of values of $\lambda$ that permits to consider the lifetime of this meta-stable vacuum sufficiently long), the term that multiplies  $\bar\psi \psi$ is indeed zero at $\varphi = \varphi_{vm}$. The massless fermion is nothing else than the usual Majorana fermion of the critical Ising Model. Increasing further $\lambda$ and reaching values much larger than $\lambda_c$, the lifetime of the metastable vacuum shortens and the effective theory based on this pseudo vacuum loses, at certain point, its validity. As a matter of fact, at strong coupling this pseudo-vacuum is going to be finally absorbed into the maximum   of the second term of $V_B$, i.e. $\sin\left(\frac{1}{3} \varphi\right)$.   It is easy to check that the situation just discussed for the case $\omega = \frac{1}{3}$  also occurs in all other cases when $(q-p)$ is an even number, possibly with several critical values of  $\lambda$ where phase transitions of the type of the Tricritical Ising Model occur.
  
Let us discuss now the second case,  $\omega = \frac{2}{3}$. The evolutation of the landscape of $V_B$ by varying $\lambda$ is shown in the sequence of pictures of Figure 12. When $\lambda$ is just switched on, the number of zeros does not change but the barriers that separate them start to lower.  If one concentrates the attention on the barriers placed in the middle of the plots, one notices that their evolution follows the patter presented by the gaussian model. Denoting by $\hat \lambda_c$ the critical value of the coupling, in the vicinity of this value the shape of the $V_B$ can be parameterized similarly to the bosonic potential of the gaussian model 
\EQ
V_B \,\sim\, (\varphi - 3\pi)^2 \, \left[(\varphi - 3 \pi)^2 + (\lambda - \hat\lambda_c) \, \right]^2 \,\,\,.
\EN  
In this case, overpassing the critical value $\hat \lambda_c$, supersymmetry still remains exact at the vacuum $\varphi^{(0)} =  3 \pi$. It is easy to see that a similar situation occurs in all other case when $(q-p)$ is an odd number, the only difference possibly being the existence of a sequence of other critical values $\lambda_c^{(n)}$ (in addition to $\hat\lambda_c$) where TIM-like phase transitions occur at other vacua. This happens if $(q - p) > 1$. If we had decide to quantize the theory by choosing as vacuum the one placed at $\varphi^{(0)} = \pi q$, the dynamics seen at this vacuum state consists, at $\lambda = \hat \lambda_c$, in a massless flow from $c = 3/2$ to $c = 1$, and then from $c =1$ to $c =0$, without a breaking of the supersymmetry.

\vspace{5mm}
\begin{figure}[h]
\hskip 25pt
\begin{minipage}[b]{.25\linewidth}
\centering\psfig{figure=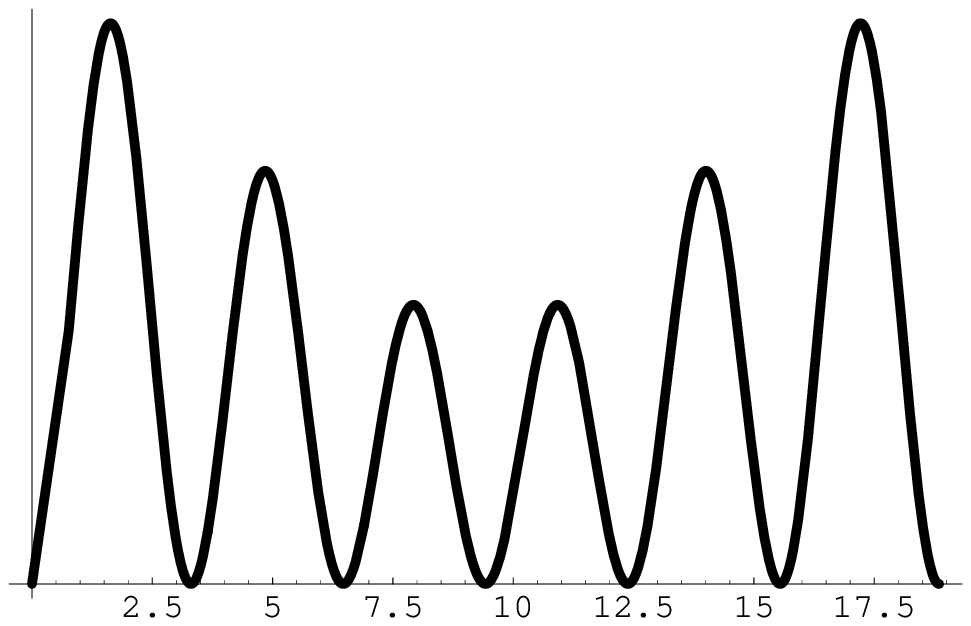,width=\linewidth}
%\caption{(a)}
%\vspace{1mm}
\begin{center}
{\bf (a)}
\end{center}
\end{minipage} \hskip 30pt
\begin{minipage}[b]{.25\linewidth}
\centering\psfig{figure=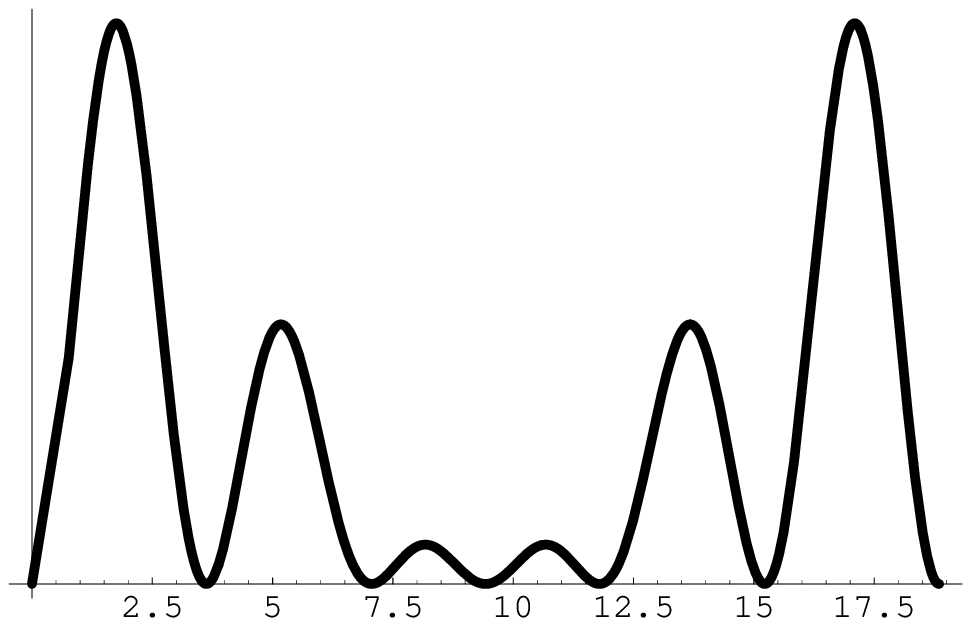,width=\linewidth}
%\caption{(b)}
%\vspace{1mm}
\begin{center}
{\bf (b)}
\end{center}
\end{minipage} \hskip 30pt
\begin{minipage}[b]{.25\linewidth}
\centering\psfig{figure=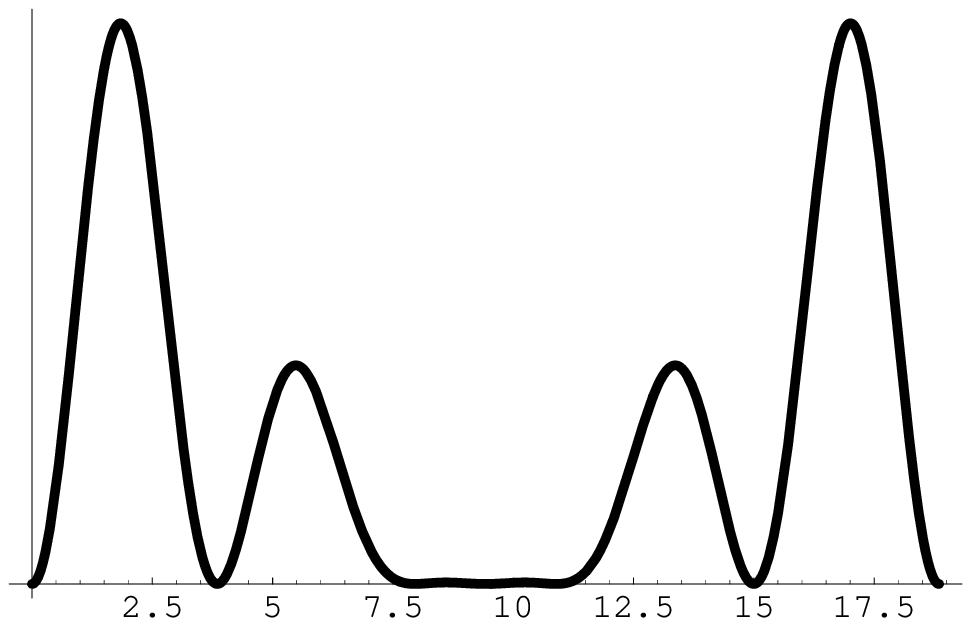,width=\linewidth}
%\caption{(a)}
%\vspace{1mm}
\begin{center}
{\bf (c)}
\end{center}
\end{minipage} \hskip 30pt
\end{figure}
\vspace{3mm}
\begin{figure}[h]
\hskip 25pt
\begin{minipage}[b]{.25\linewidth}
\centering\psfig{figure=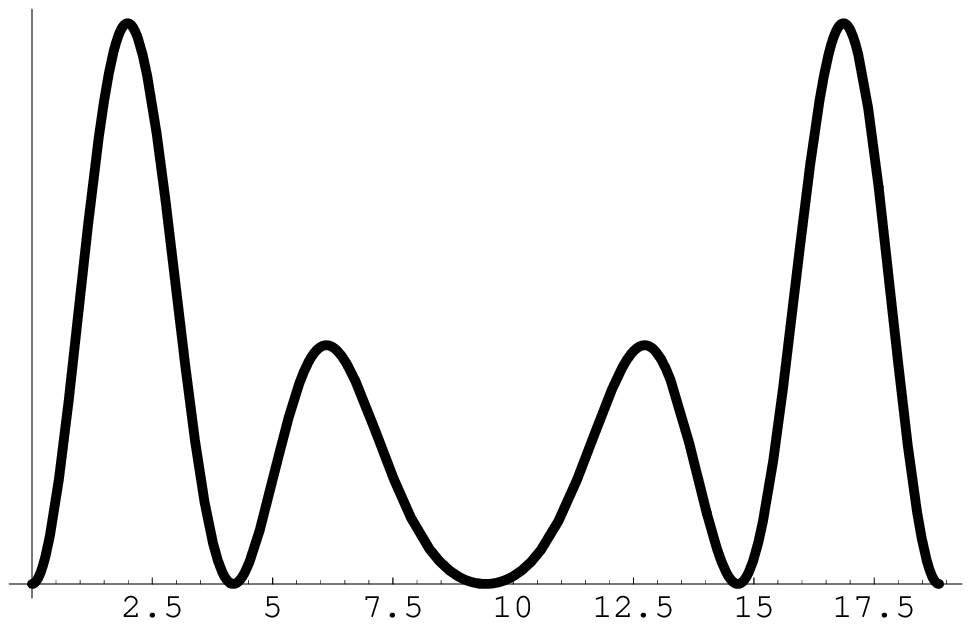,width=\linewidth}
\begin{center}
{\bf (d)}
\end{center}
\end{minipage} \hskip 30pt
\begin{minipage}[b]{.25\linewidth}
\centering\psfig{figure=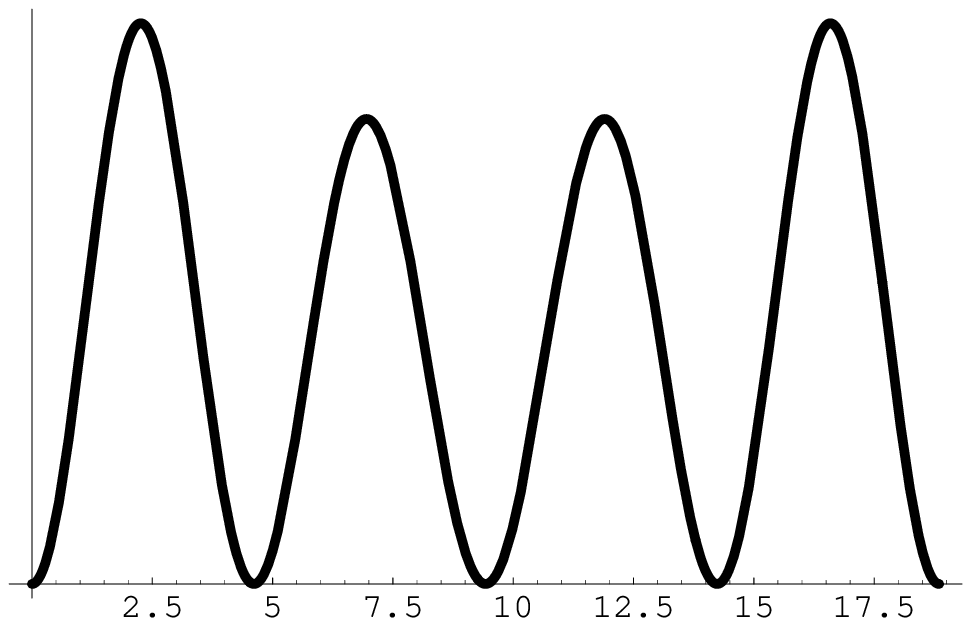,width=\linewidth}
%\caption{(b)}
%\vspace{1mm}
\begin{center}
{\bf (e)}
\end{center}
\end{minipage} \hskip 30pt
\begin{minipage}[b]{.25\linewidth}
\centering\psfig{figure=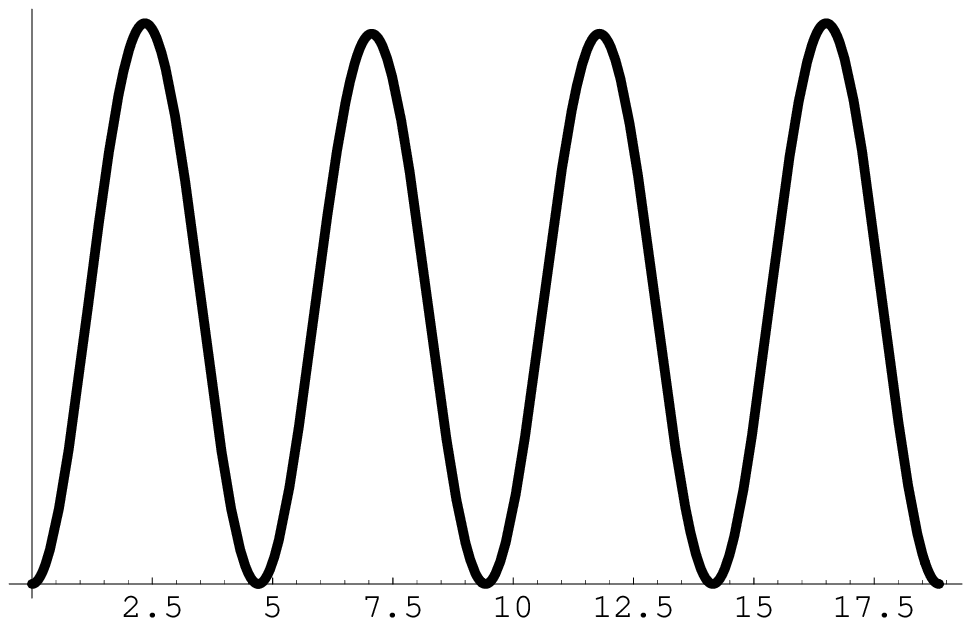,width=\linewidth}
%\caption{(a)}
%\vspace{1mm}
\begin{center}
{\bf (f)}
\end{center}
\end{minipage} 
\caption{{\em $V_B$ for $\omega = 2/3$ by varying  $\lambda$: (a) small value of $\lambda $ (b) $\lambda < \hat\lambda_c$; (c) $\lambda = \hat\lambda_c$; (d) $\lambda > \hat\lambda_c$; (e) $\lambda \gg \hat\lambda_c$; (f) $\lambda \rightarrow \infty$ . }}
\end{figure}

\resection{Conclusions}

In this paper we have analysed, through the study of a particular model, the breaking of integrability of a supersymmetric theory. On the contrary to purely bosonic theories, this does not lead to a confinement of the kink exitations. The absence of this phenomenon is due precisely to the unbroken supersymmetry invariance of the theory. However, by varying the coupling constant, the spectrum changes. 

At weak coupling, there is no longer the degeneracy of the original kinks: the kinks split into a sequence of {\em long} and {\em short} kinks. Since they are still BPS states, their mass can be computed in terms of their topological charge ${\cal Z}$. An interesting open problem consists of  comparing the values obtained by this approach with the masses computed in terms of the Form Factor Perturbation Theory. In order to do so, one needs to generalize the supersymmetric Form Factors analysed in \cite{GMsusy} to theories with kinks. 

At finite values of the coupling, there exist critical points where phase transitions take place in the system: at these critical values, pairs of kinks become massless and then disappear from the set of asymptotic states. 

Among the phase transitions that take place in the theory, some of them realise a spontaneously supersymmetry breaking around certain meta-stable vacua. This happens if $(q - p) > 1$, 
where $\omega = p/q$ is the frequency of the perturbing superfield. A quantization of the theory around the metastable vacua leads, nearby the critical values of the coupling, to the effective dynamics that  describes the spontaneously supersymmetry breaking of the Tricritical Ising model. This consists of a massless flow from this model to the critical Ising model, going from the 
short to the large distance scales. In this flow, the scalar particle is massive while the fermion is the massless Mayorana particle of the Ising model, i.e. the goldstino of the supersymmetry breaking. 

If $(q -p)$ is an odd number, the system also present a phase transition at $\hat\lambda_c$ 
which does not break supersymmetry at the vacuum in the middle of the interval $(0, 2 \pi q)$ where it occurs. In this case, the effective dynamics seen at this vacuum consists of the class of universality of the gaussian model, which presents massive and supersymmetric excitations either before or after the critical point, becoming massless just at $\lambda = \hat\lambda_c$: for $\lambda < \hat \lambda_c$ they appear as kinks, for $\lambda > \hat \lambda_c$, they are instead ordinary particles. 
 
\vspace{15mm}
\noindent
{\em Acknowledgments}. I would like to thank G. Delfino and P. Fendley for useful discussions. 
The work is partially supported by the ESF grant INSTANS and by the MUR grant "Quantum Field Theory and Statistical Mechanics in Low Dimensions".

\end{document}